\documentclass[a4paper,fleqn,usenatbib]{mnras}

\usepackage{color}
\usepackage{epsfig}
\usepackage{graphicx}
\usepackage{array}
\usepackage{color}
\usepackage{comment}
\usepackage[utf8]{inputenc}  
\usepackage{amsmath,breqn}  
\usepackage{natbib}
\bibpunct{(}{)}{;}{a}{,}{,}

\usepackage{amssymb}
\usepackage{float}
\usepackage{tabularx}
\usepackage{footnote}
\usepackage{booktabs}
\usepackage[T1]{fontenc}
\usepackage{url}
\usepackage{cleveref}
\usepackage{aas_macros}

\title[Eclipsing DLAs in SDSS DR12]{Eclipsing damped Ly$\alpha$ systems in the Sloan Digital Sky Survey Data Release 12\thanks{Full version of Table\,1 and 2 are available at CDS via anonymous ftp to cdsarc.u-strasbg.fr (130.79.128.5) or via http://cdsarc.u-strasbg.fr/viz-bin/qcat?J/MNRAS}}

\author[H. Fathivavsari et al.]{
 H.~Fathivavsari,$^{1}$\thanks{E-mail: h.fathie@gmail.com} P.~Petitjean$^{2}$, N.~Jamialahmadi$^{1}$, H.~G.~Khosroshahi$^{1}$,
 \newauthor 
 H.~Rahmani$^{3}$,
 H.~Finley$^{4}$,
 P.~Noterdaeme$^{2}$,
 I.~P\^aris$^{5}$,
 R.~Srianand$^{6}$\\
$^{1}$School of Astronomy, Institute for Research in Fundamental Sciences (IPM), P. O. Box 19395-5531, Tehran, Iran\\
$^{2}$Institut d'Astrophysique de Paris, Universit\'e Paris 6-CNRS, UMR7095, 98bis Boulevard Arago, 75014 Paris, France \\
$^{3}$GEPI, Observatoire de Paris, PSL Research University, CNRS, Place Jules Janssen, F-92190 Meudon, France\\
$^{4}$Institut de Recherche en Astrophysique et Plan\'etologie (IRAP), Universit\'e de Toulouse, CNRS, UPS, F-31400 Toulouse, France\\
$^{5}$Aix Marseille Univ, CNRS, LAM, Laboratoire d’Astrophysique de Marseille, Marseille, France\\
$^{6}$Inter-University Centre for Astronomy and Astrophysics, Post
Bag 4, Ganeshkhind, 411 007, Pune, India\\
}

\date{Accepted 000. Received 000; in original form 000}

\pubyear{2018}

\begin{document}
\label{firstpage}
\pagerange{\pageref{firstpage}--\pageref{lastpage}}
\maketitle

\begin{abstract}

We present the results of our automatic search for proximate damped Ly$\alpha$ absorption (PDLA) systems in the quasar spectra from the Sloan Digital Sky Survey Data Release 12. We constrain our search to those PDLAs lying within 1500\,km\,s$^{-1}$ from the quasar to make sure that the broad DLA absorption trough masks most of the strong Ly$\alpha$ emission from the broad line region (BLR) of the quasar.  When the Ly$\alpha$ emission from the BLR is blocked by these so-called \emph{eclipsing} DLAs, narrow Ly$\alpha$ emission from the host galaxy could be revealed as a narrow emission line (NEL) in the DLA trough.  We define a statistical sample of 399 eclipsing DLAs with log\,$N$(H\,{\sc i})\,$\ge$\,21.10. We divide our statistical sample into three subsamples based on the strength of the NEL detected in the DLA trough. By studying the stacked spectra of these subsamples, we found that absorption from high ionization species are stronger in DLAs with stronger NEL in their absorption core. Moreover, absorption from the excited states of species like Si\,{\sc ii} are also stronger in DLAs with stronger NEL. We also found no correlation between the luminosity of the Ly$\alpha$ NEL and the quasar luminosity. These observations are consistent with a scenario in which the DLAs with stronger NEL are denser and physically closer to the quasar. We propose that these eclipsing DLAs could be the product of the interaction between infalling and outflowing gas. High resolution spectroscopic observation would be needed to shed some light on the nature of these eclipsing DLAs.

\end{abstract}




\section{Introduction}

Models of galaxy formation and evolution require gas inflow and outflow as basic ingredients to reproduce the observed properties of galaxies   \citep{1998A&A...331L...1S, 2003MNRAS.345..349B,2012MNRAS.421.2809V,2016ApJ...824L...5M}. The infall of gas on to the galaxy occurs preferentially through cold streams along the filaments of the cosmic web  \citep{2005MNRAS.363....2K}. Although the existence of this so-called cold mode accretion has been firmly established by simulations, little direct observational evidence is available for its existence \citep{2008ApJ...681..856R, 2010ApJ...717..289S,2010Natur.467..811C,2011ApJ...743...95G}. Instead, outflows driven by supernovae and/or active galactic nuclei (AGN) are ubiquitously observed as blue-shifted absorption features in quasar and galaxy spectra at any redshift \citep{2002MNRAS.336..753S,2015ApJ...811..149C,2015MNRAS.452.2712W, 2016ApJ...833...39S, 2016ApJ...818...28S,2017A&A...605A.118F}.

It would be logical to think that pristine infalling gas should be detected as redshifted absorption features in the spectra of galaxies and quasars \citep{2011MNRAS.413L..51K, 2011ApJ...735L...1S}. However, several reasons have been proposed as to why these cold flows should be spectroscopically difficult to detect.
For example, if the velocity of the infalling gas is not large enough then deep and broad (in velocity space) absorption features from the interstellar medium of the galaxy and/or from enriched outflowing gas would most probably contaminate and drown the absorption signals from the chemically young infalling gas \citep{2011MNRAS.413L..51K, 2012ApJ...747L..26R}. However, in cases where the line of sight solely passes through the cold flows, these contaminations could be avoided.

Simulations have shown that cold streams could be detected as low metallicity ($Z/Z_{\odot}$\,$\sim$\,10$^{-3}$\,$-$\,10$^{-2}$) Lyman Limit Systems (LLSs) with neutral hydrogen column density of $\sim$\,10$^{17}$\,cm$^{-2}$     \citep{2011MNRAS.412L.118F,2011MNRAS.418.1796F, 2012MNRAS.423.2991V, 2013ApJ...765...89S}. Motivated by these simulations, many studies have been devoted to characterise the physical properties of LLSs and probe their possible connection with the cold flows \citep{2013ApJ...770..138L, 2015ApJ...812...58C,2015ApJS..221....2P, 2016ApJ...833..283L, 2016ApJ...833..270G,2016ApJ...831...95W,2018MNRAS.474..254R}. Using the Cosmic Origin Spectrograph on board the \emph{Hubble Space Telescope}, \citet{2016ApJ...831...95W} studied 55 H\,{\sc i}-selected LLSs at $z$\,$\lesssim$\,1 and found evidence for a bimodal distribution of metallicity with two well-separated peaks at [X/H]\,$\simeq$\,$-$1.87 and $-$0.32. Moreover, with the aim of exploring the evolution of LLS metallicities with redshift, \citet{2016ApJ...833..283L} studied 31 H\,{\sc i}-selected LLSs at $z$\,$>$\,2, observed with the HIRES spectrograph on the Keck telescope, and found that the LLS metallicity distribution function evolves with redshift, changing from a bimodal distribution at $z$\,$<$\,1 to a unimodal distribution (with a peak at [X/H]\,$\simeq$\,$-$2) at $z$\,$>$\,2. These authors suggest that the low-metallicity branch of the distribution may trace chemically young cold flows and the high-metallicity LLSs probably represent enriched outflowing gas.

It could be possible that enriched AGN- and/or supernova-driven outflowing gas, on its way to the outer region of the galaxy, may collide with metal-poor ($Z/Z_{\odot}$\,$\sim$\,10$^{-3}$\,$-$\,10$^{-2}$) and/or primordial ($Z/Z_{\odot}$\,$\lesssim$\,10$^{-3}$) infalling gas. If the two gas collide, they get mixed, shocked and compressed to a high density \citep{2014MNRAS.443.2018N}.
If the density gets high enough, then the compressed gas could give rise to a DLA when seen in absorption against a background source such as a quasar. Since the DLA and the background quasar are both located almost at the same redshift, the DLA can act as a natural coronagraph, blocking the quasar blazing radiation in Ly$\alpha$. This would then allow us, depending on the dimension of these so-called \emph{eclipsing} DLAs, to detect fainter emission from the star-forming regions in the host galaxy and/or from the narrow line region (NLR) of the AGN  \citep{2009ApJ...693L..49H,2013A&A...558A.111F, 2015MNRAS.454..876F,2016MNRAS.461.1816F}. With a nominal H\,{\sc i} density of 10\,cm$^{-3}$ \citep{2009ApJ...690.1558P, 2016MNRAS.461.1816F}, a DLA with H\,{\sc i} column density of 10$^{21}$\,-\,10$^{22}$\,cm$^{-2}$ would have a characteristic size (i.e. $l$\,$\sim$\,30\,-\,300\,pc) smaller than the NLR of the quasar \citep{2002ApJ...574L.105B, 2006ApJ...646...16Y}.
Consequently, the DLA absorber would not be able to fully block the emission from the NLR and/or star-forming regions, and the leaked emission from these regions could be detected as a narrow Ly$\alpha$ emission in the DLA trough.
On the other hand, if the covering factor of an eclipsing DLA is 100 per cent  then no Ly$\alpha$ narrow emission would be detected in the DLA trough. Observing a DLA along the line of sight to a background galaxy, \citet{2015ApJ...812L..27C} could put a constraint on the size of the DLA by considering the fact that the DLA almost fully covers the extended background galaxy.
In extreme cases where the hydrogen density is so high (i.e. $n_{\rm HI}$\,>\,1000\,cm$^{-3}$) and the DLA size is smaller than the size of the BLR of the quasar, the leaked broad Ly$\alpha$ emission from the BLR could fully fill the DLA trough resulting in the formation of a \emph{ghostly} DLA in the quasar spectrum \citep{2017MNRAS.466L..58F}. These DLAs are called ghostly as they reveal no DLA absorption profile in their spectra. The characterization of these kind of systems is extremely important to understand the details of how the gas is accreted on to and/or thrown out by the AGN.

With the aim of statistically studying the properties of eclipsing and ghostly DLAs, we search the SDSS-III Baryon Oscillation Spectroscopic Survey Data Release 12 \citep[BOSS;][]{2013AJ....145...10D} for such absorbers. This is similar to what was done in  \citet{ 2013A&A...558A.111F} but using a much larger sample. In the current work we also employ a more robust searching algorithm to find the DLA absorbers. In this paper, we focus on understanding the physical properties of eclipsing DLAs while characterization of ghostly DLAs will be presented in another paper.

\begin{figure*}
\centering
\begin{tabular}{c}
\includegraphics[bb=46 436 559 577, clip=,width=0.90\hsize]{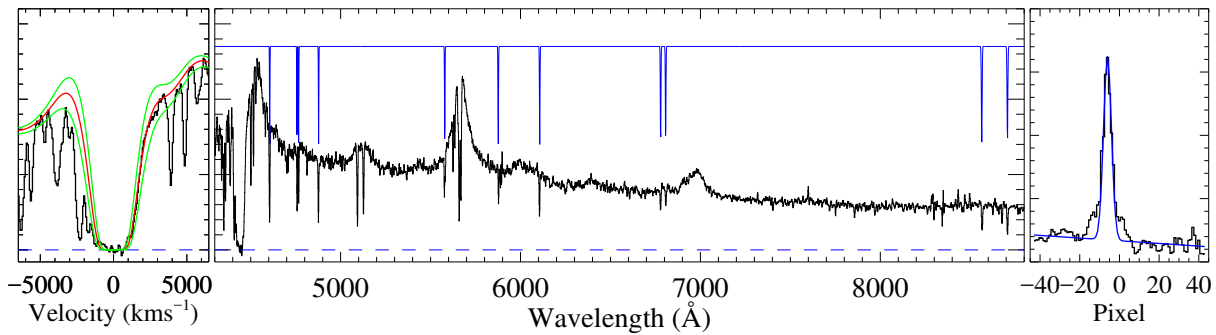}
\end{tabular}
\caption{Example of the SDSS spectrum of J015857.66+320233.3 with a DLA absorption at $\lambda_{\rm obs}\,\sim$\,4426\,\textup{\AA}. Left panel: example of a Voigt profile fit to the DLA absorption profile. The two green curves show the uncertainty of $\pm$\,0.10\,dex on the column density. Middle panel: the low ionization lines template used to find candidate metal absorbers is overplotted in blue on the observed spectrum. Right panel: example of the cross-correlation function. The wavelength of the pixel at which the maximum occurs gives us the absorption redshift of the absorber.}
 \label{ccf}
\end{figure*}

\section{Method} \label{sectmethod}

DLA candidates in the SDSS spectra are generally identified by their large Ly$\alpha$ absorption trough. The current automatic approaches to find DLAs require the flux at the bottom of the DLA to be at zero level \citep{2004PASP..116..622P,2009A&A...505.1087N}. However, when a narrow Ly$\alpha$ emission is present in a DLA trough, it may look like as if we have two consecutive DLAs. In these cases, the current approaches may wrongly identify these two troughs as two individual DLA systems; or even they could detect only one of the troughs and skip the other one if the SNR is low and/or the blue wing of the DLA is contaminated by forest absorption. Moreover, if the narrow emission is rather strong, the whole DLA could skip detection as the presence of the emission in the DLA trough could significantly affect the correlation function.

In this paper, we follow a rather different approach to search for eclipsing DLA candidates in the BOSS spectra. Our searching procedure is based on the detection of metal absorption lines in quasar spectra. We take into account only quasars with the emission redshift higher than $z_{\rm em}$\,=\,2.0 as this is the minimum redshift for which the Ly${\alpha}$ transition spectral region falls on the observed spectral window.
Quasars with Broad Absorption Lines (BALs) are excluded from our sample as their blending with the DLA absorption could contaminate the H\,{\sc i} column density measurements. In the BOSS DR12 catalog of \citet{2017A&A...597A..79P}, quasars with BALs are recognized by their non-zero balnicity index \citep{1991ApJ...373...23W}. Moreover, since our searching procedure strongly depends on the automatic recognition of the DLA absorption profile, we therefore constrain our search to spectra with the continuum-to-noise ratio (CNR) above 4.0. The CNR is measured over the spectral region between 1410 to 1510\,\textup{\AA} in the rest frame. Applying these criteria on the BOSS spectra returns 149387 quasars. We call this sample of quasars, the $S^{1}_{QSO}$ sample.

\subsection{Search for metal absorption lines} \label{searchingmethod}

We use a Spearman correlation analysis \citep{2014Ap&SS.353..347F} to find the candidate metal absorption line systems. To this end, we construct an absorption template made up of 12 strong metal absorption lines commonly detected in DLAs, i.e., O\,{\sc i}$\lambda$1302, C\,{\sc ii}$\lambda$1334, Si\,{\sc ii}$\lambda$1260, 1304, 1526, Al\,{\sc ii}$\lambda$1670, Al\,{\sc iii}$\lambda$1854, 1862, Fe\,{\sc ii}$\lambda$1608, 2344, 2382, 2600 (see Fig.\,\ref{ccf}). We choose only those metal lines located redwards of the quasar's Ly$\alpha$ emission, in order to avoid contamination by the forest absorption lines. Since the width of the DLA trough is $\sim$\,2000\,km\,s$^{-1}$ at log\,$N$(H\,{\sc i})\,$\sim$\,21.0, therefore, the NEL from the quasar host galaxy and/or NLR falls inside the DLA trough \emph{only if} the redshift of the DLA is within $\sim$\,1000\,km\,s$^{-1}$ of the quasar emission redshift (i.e. $\mid$$\Delta$V$\mid$\,$\le$\,1000\,km\,s$^{-1}$). In our search for metal absorbers, we will adopt the conservative value of $\mid$$\Delta$V$\mid$\,$\le$\,1500\,km\,s$^{-1}$ in order to take into account the error in quasar emission redshift estimation \citep{2010MNRAS.405.2302H}.

At each pixel {\sl j} in the spectrum, corresponding to the redshift $z_{j}$ lying within $\pm$\,1500\,km\,s$^{-1}$ of the quasar's redshift, the metal absorption template is successively correlated with the observed spectrum, and the corresponding correlation coefficient is calculated. This exercise gives us one correlation function for each quasar spectrum. If the correlation function has a maximum with high significance ($\ge$\,4\,$\sigma$), the system is recorded, and the redshift at which the maximum occurs is taken as the absorption redshift of the candidate system. Employing this constraint on the $S^{1}_{QSO}$ sample leaves us with 45040 systems, many of which are false-positive detections. We call this new sample, the $S^{2}_{QSO}$ sample. We checked that even if only 2 unrelated intervening absorption lines (detected above 4\,$\sigma$) are coincidentally located at the same position of any two absorption lines of the template, the correlation function can peak with $\ge$\,4\,$\sigma$ significance. Also, the presence of a sharp spike in the spectrum, which is usually caused by the imperfect sky emission subtraction, can result in false-positive detections.

In order to exclude from the $S^{2}_{QSO}$ sample, the false-positive detections caused by the intervening absorption lines and sharp spikes, we calculate the equivalent width ($W$) and its uncertainty ($\sigma_{W}$) for the 12 metal transitions included in the template. We then keep only those candidate systems for which at least 3 absorption lines are detected above 3\,$\sigma$ (i.e. $W$\,$\ge$\,3\,$\sigma_{W}$). Employing this algorithm on the $S^{2}_{QSO}$ sample returns 10224 systems. We call this new sample, the $S^{3}_{QSO}$ sample. Since the candidate systems are found through the cross-correlation of a metal absorption line template with the observed spectra, in principle, the $S^{3}_{QSO}$ sample could contain Lyman Limit systems (LLS), sub-DLAs, eclipsing DLAs with and without a narrow Ly$\alpha$ emission in their troughs, ghostly DLAs and false-positive detections. Here, we try to distinguish between eclipsing DLAs (with and without NEL) and other detections.

\begin{figure*}
\centering
\begin{tabular}{c}
\includegraphics[bb=40 362 554 577, clip=,width=0.90\hsize]{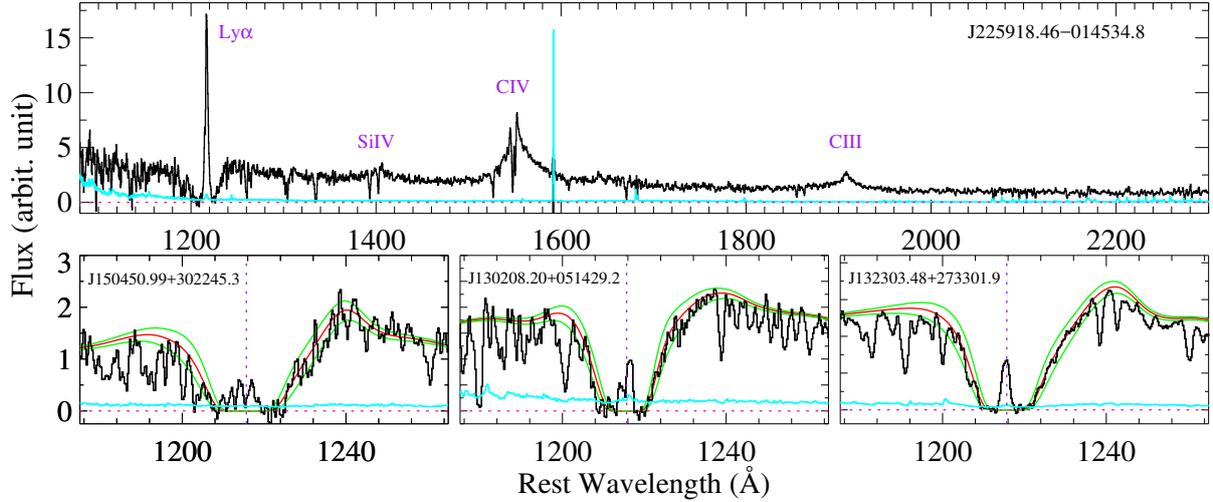}
\end{tabular}
\caption{Four sample quasar spectra with narrow Ly$\alpha$ emission in their DLA troughs. In the lower panels, only the DLA absorption regions are shown. Here, the red curves are the DLA Voigt profile fit with the two green curves showing the profile variations corresponding to a 0.1\,dex in the H\,{\sc i} column density measurement. In each panel, the error spectrum is shown in cyan. As illustrated in these plots, the narrow Ly$\alpha$ emission are observed to exhibit one, two, or multiple components.}
 \label{example_plots}
\end{figure*}

For each candidate system in the $S^{3}_{QSO}$ sample, we perform a cross-correlation of the observed spectrum with a series of synthetic DLA absorption profiles, corresponding to column densities in the range 20.5\,$\le$\,log\,$N$(H\,{\sc i})\,$\le$\,22.5, with the redshift fixed to that obtained above. The lower limit on the H\,{\sc i} column density is set to 20.5 as it gets more difficult to reliably detect NEL in the noisy troughs of DLAs with lower $N$(H\,{\sc i}). When a DLA absorption with log\,$N$(H\,{\sc i})\,$\ge$\,20.5 is present in the spectrum, the correlation coefficient is larger than 0.7. Our simulations have shown that even if a narrow Ly$\alpha$ emission is present at the bottom of the DLA, the correlation coefficient still stays above 0.7. Employing this procedure on the $S^{3}_{QSO}$ sample returns 605 systems. We call this new sample, the $S^{4}_{QSO}$ sample.  In principle, the $S^{4}_{QSO}$ sample comprises both the eclipsing DLAs with and without a NEL in their troughs. We present, in Fig.\,\ref{example_plots}, examples of four quasar spectra with eclipsing DLAs. In the lower panels of this figure, only the DLA absorption spectral regions are shown.

\subsection{The completeness of the sample} \label{completeness}

Since our search for the eclipsing DLA absorbers was primarily based on the cross-correlation of the observed spectra with a metal template, those systems with very weak metal absorption lines and/or low SNR could skip detection. However, those eclipsing DLAs could still be easily identified by their strong DLA absorption feature. Here, we try to assess the fraction of the DLAs (without a NEL) missed by our metal absorption template technique. For this purpose, we follow an approach similar to the \citet{2009A&A...505.1087N} technique with some modification.

At each pixel {\sl j} in the spectrum, corresponding to the redshift $z_{j}$ lying within $\pm$\,1500\,km\,s$^{-1}$ of the quasar's redshift, we perform a cross-correlation of the observed spectrum with a series of synthetic DLA absorption profiles.
When the correlation coefficient is $\ge$\,0.7 with high significance ($\ge$\,4\,$\sigma$), the system is recorded. Employing this algorithm on the $S^{1}_{QSO}$ sample triggered 1966 systems with 20.5\,$\le$\,log\,$N$(H\,{\sc i})\,$\le$\,21.9.

\begin{figure}
\centering
\begin{tabular}{c}
\includegraphics[bb=108 392 458 577, clip=,width=0.90\hsize]{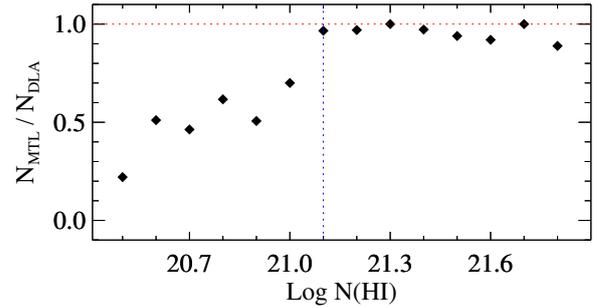}
\end{tabular}
\caption{The ratio of the number of DLAs found by metal template method to the number of DLAs found through DLA template technique as a function of H\,{\sc i} column density. The ratio reaches $\sim$\,1.0 at H\,{\sc i} column density of $\sim$\,21.10, implying that the sample is complete above this column density (see the text).}
 \label{N1N2Ratio}
\end{figure}

Figure\,\ref{N1N2Ratio} shows the ratio of the number of DLA systems found by metal template technique (N$_{\rm MTL}$) to the number of systems found through DLA template approach (N$_{\rm DLA}$) as a function of H\,{\sc i} column density. As shown in this figure, the ratio reaches $\sim$\,1 at log\,$N$(H\,{\sc i})\,$\sim$\,21.10, meaning that our sample is complete above this column density. The divergence between the two approaches at lower H\,{\sc i} column densities mainly stems from the fact that the metal template technique is susceptible to miss systems with weak metal absorption lines and/or low SNR. We, therefore, only use the properties of the eclipsing DLAs with log\,$N$(H\,{\sc i})\,$\ge$\,21.10 when we statistically look for differences between the eclipsing DLAs with and without narrow Ly$\alpha$ emission.

\begin{figure}
\centering
\begin{tabular}{ll}
\includegraphics[bb=117 395 464 559, clip=,width=0.90\hsize]{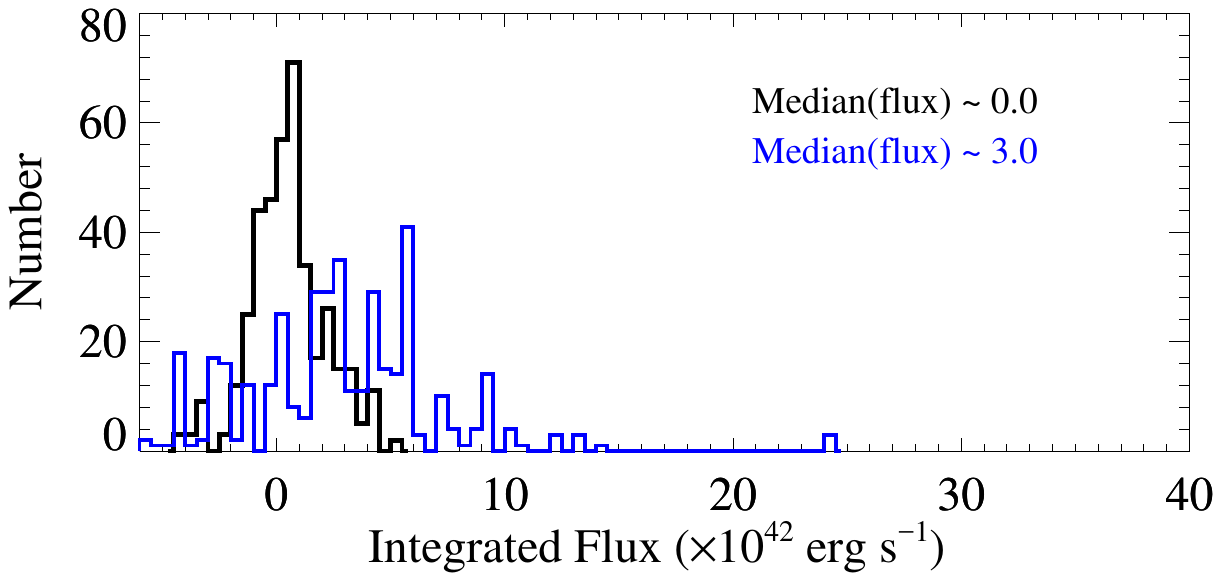}\\
\includegraphics[bb=117 395 464 560, clip=,width=0.90\hsize]{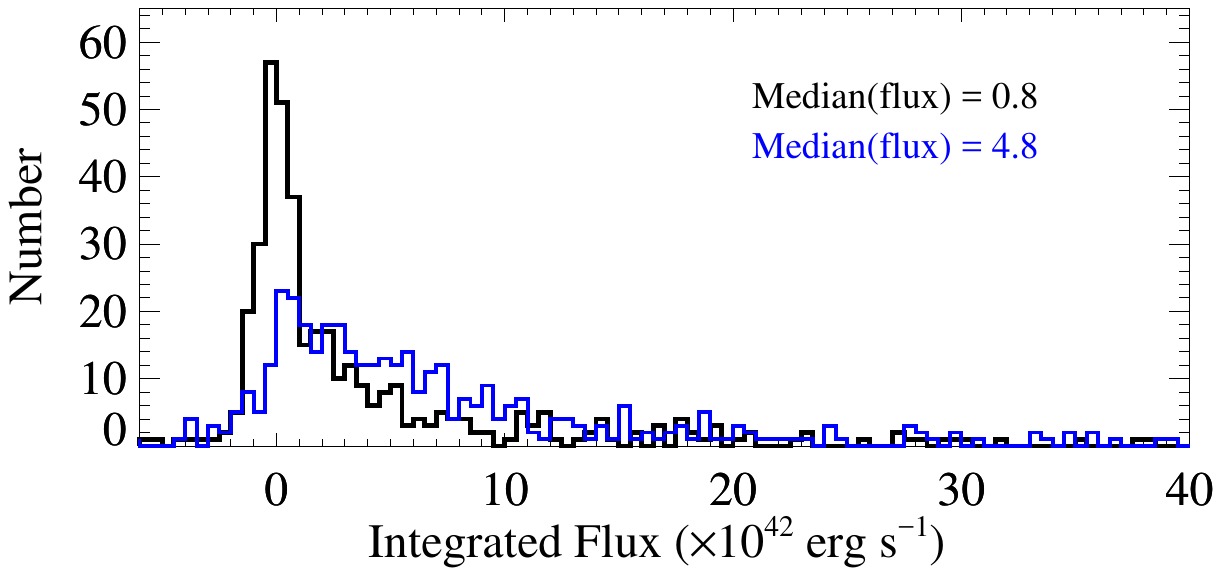}
\end{tabular}
\caption{Histograms of the integrated flux in the DLA troughs before (blue histogram) and after (black histogram) the zero-level correction. The median of the integrated flux before and after correction are also shown. The upper panel shows the results for the intervening DLAs while the lower panel is for the eclipsing DLAs from our statistical sample.}
 \label{SLHist}
\end{figure}

\subsection{Zero-level correction and NEL detection}
The flux zero-level in the BOSS spectra is slightly above zero, especially in the blue part of the spectra where we have the DLA absorption \citep{2018MNRAS.476..210J}. These residual flux could falsely be interpreted as the detection of a NEL in the DLA trough.
Therefore, we should first correct for this effect before determining whether any emission is detected in the DLA trough. To do this, for each spectrum, we define the median flux of the pixels in the DLA trough as the new zero-flux level and then adjust the DLA trough accordingly.  Here, the DLA trough is defined as the region over which the flux is below 1 per cent of the continuum. We use synthetic DLA absorption to determine the number of pixels (in the DLA trough) for which the median is calculated.

We checked the applicability of this procedure by applying it on 399 intervening DLAs from the \citet{2012A&A...547L...1N} catalog. We chose only those intervening DLAs that were located at $\ge$\,10000\,km\,s$^{-1}$ from the quasar, and had absorption redshift and H\,{\sc i} column densities similar to those from our statistical sample.
We show in the upper panel of Fig.\,\ref{SLHist}, the histogram of the integrated flux (IF; in erg\,s$^{-1}$\,cm$^{-2}$) in the intervening DLA troughs before (blue histogram) and after (black histogram) the zero-level correction. As shown in this figure, the median of the IF decreases from $\sim$\,3.0 to $\sim$\,0.0 after applying the correction. We then apply this procedure on our eclipsing DLA sample to correct for the zero flux level in the DLA trough. The result is shown in the lower panel of Fig.\,\ref{SLHist}. As shown in this figure, the median of the IF decreases from $\sim$\,4.8 to $\sim$\,0.8.

Sometimes this procedure underestimates or overestimates the zero flux level, especially when a Ly$\alpha$ NEL is present in the DLA trough.
We visually inspect all our systems and found that for 40 eclipsing DLAs,  the zero level was not correctly estimated by our automatic procedure. We therefore tried to visually re-adjust the zero level in these DLAs.  The visual inspection was performed independently by the first two authors of the paper.
From the visual inspection, we confirm the detection of NEL in the troughs of 155 eclipsing DLAs from our statistical sample.  We note that the possibility that these NELs could be the [O\,{\sc ii}] and/or [O\,{\sc iii}] emission from some intervening low redshift galaxies is very small \citep{2010MNRAS.403..906N, 2017MNRAS.471.1910J}. However, we still checked this possibility for the strongest Ly$\alpha$ NELs in our sample and found that none of them could be attributed to the [O\,{\sc ii}] and/or [O\,{\sc iii}] emission from intervening galaxies.

\subsection{Measuring accurate H\,{\sc i} column densities}

\begin{figure}
\centering
\begin{tabular}{c}
\includegraphics[bb=114 398 458 649, clip=,width=0.90\hsize]{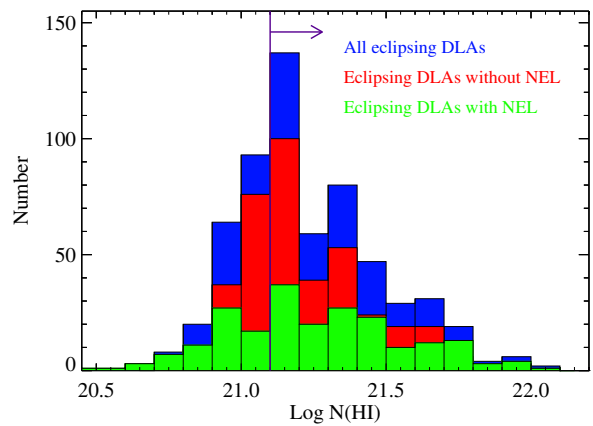}
\end{tabular}
\caption{H\,{\sc i} column density distributions for the eclipsing DLAs in the SDSS-III DR12. The green and red indicate DLAs with and without narrow Ly$\alpha$ emission, respectively. The blue presents all DLAs. The sample is complete above log\,$N$(H\,{\sc i})\,=\,21.10 as indicated by the purple arrow.}
 \label{NHI_dist}
\end{figure}

Before measuring the H\,{\sc i} column densities, we need to normalise the spectra. Since the broad Ly$\alpha$ emission of the quasars is extinguished by the eclipsing DLAs, the principle component analysis \citep[PCA;][]{2011A&A...530A..50P} prediction of the quasars' continua around the Ly$\alpha$ spectral region is used to normalize the spectra. We measure the H\,{\sc i} column density of the DLAs by fitting a Voigt profile function on the DLA absorption profile with the redshift fixed to that of the associated metal absorption lines.  The resultant H\,{\sc i} column densities of the eclipsing DLAs in our sample are listed in Table\,\ref{bigtable1} and their distribution is presented in Fig~\ref{NHI_dist}. Errors on the measured column densities are on the order of 0.1~dex. We note that our automatic estimation of the H\,{\sc i} column density of the eclipsing DLAs (as described in section\,\ref{sectmethod}) does not differ by more than 0.1\,dex from the accurate measurements. As shown in Fig.\,\ref{NHI_dist}, the maximum in the H\,{\sc i} column density distribution occurs at log\,$N$(H\,{\sc i})\,=\,21.10. This further justifies our cut at this column density when defining our statistical sample in section\,\ref{completeness}.

\section{results}
In this section,
we only take into account the eclipsing DLAs from the statistical sample (see section\,\ref{completeness}). We recall that our statistical sample comprises 399 eclipsing DLAs with log\,$N$(H\,{\sc i})\,$\ge$\,21.10.

\subsection{Eclipsing DLAs clustering around quasars}

In order to probe the possible clustering of DLAs around quasars, we check whether the incidence of our eclipsing DLAs is higher than that of intervening DLAs. To this end, we use the column density distribution of intervening DLAs (i.e. $f(N_{\rm HI},\,X)$) given in \citet{2014A&A...566A..24N} to calculate the anticipated number of such DLAs within 1000\,km\,s$^{-1}$ of the quasar redshift (i.e. $N_{\rm DLA}^{a}$). We perform the following integration to estimate $N_{\rm DLA}^{a}$,
\begin{equation}
N_{\rm DLA}^{a} = \int\Delta\,X\,f(N_{\rm HI},\,X)\,dN_{\rm HI}.
\end{equation}
The integration is performed over the H\,{\sc i} column density range of 21.10\,$\le$\,$N_{\rm HI}$\,$\le$\,22.30 as our statistical sample is complete over this column density range. The absorption path $X$ is defined as
\begin{equation}
X\,=\,\int_{0}^{z}\frac{H_{0}}{H(z)}\,(1\,+\,z)^{2}\,dz \,\, ,
\end{equation}
where $H_{0}$ is the present value of Hubble's constant and $H(z)$ is its value at any redshift $z$. The absorption path for each quasar is calculated over the redshift range corresponding to $-$1000\,$\le$\,$\Delta$V\,$\le$\,0\,km\,s$^{-1}$ of the quasar emission redshift. The total absorption path of 149387 quasars in the $S^{1}_{QSO}$ sample is $\Delta\,X$\,$\approx$\,6533. From the equation\,1, the expected number of intervening DLAs with log\,$N$(H\,{\sc i})\,$\ge$\,21.10 located within 1000\,km\,s$^{-1}$ of the quasar emission redshift is $N_{\rm DLA}^{a}$\,=\,61. Our statistical sample, on the other hand, contains 135 such DLAs which is $\sim$\,2.2 times more than what is expected from the column density distribution of intervening DLAs. This value is a lower limit due to the uncertainty in the quasar emission redshift estimation.
This is consistent with what \citet{2013A&A...558A.111F} found for a smaller sample of eclipsing DLAs in SDSS DR9. Moreover, \citet{2010MNRAS.406.1435E} also found an overdensity of $\sim$\,2$-$4 times in the incidence of proximate DLAs (located within 3000\,km\,s$^{-1}$ of the quasar) compared to that of intervening DLAs. One could attribute this observed overdensity to the clustering of proximate and/or eclipsing DLAs around quasars as an overdensity of neutral gas within $\sim$\,10\,Mpc of the quasar has been reported \citep{2007MNRAS.377..657G,2007ApJS..171...29P,2013JCAP...05..018F}.

\subsection{Equivalent width distribution}

\begin{figure}
\centering
\begin{tabular}{c}
\includegraphics[bb=100 388 458 614, clip=,width=0.90\hsize]{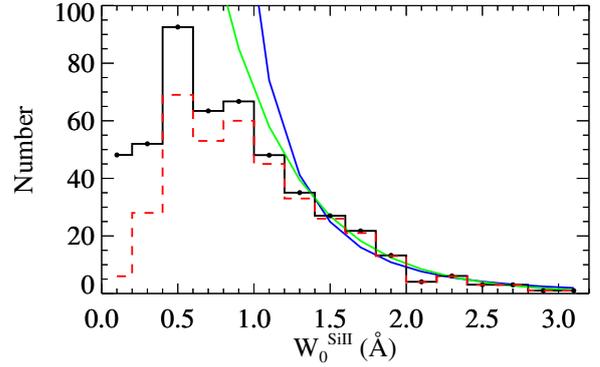}
\end{tabular}
\caption{Red dashed histogram shows the rest-frame Si\,{\sc ii}\,$\lambda$1526 EW distribution for the eclipsing DLAs from our statistical sample. The black histogram represents the distribution after applying the correction for the possible miss of systems using the total redshift path function, $g(W)$. The green and blue curves show the exponential and power law fits, respectively.}
 \label{si2ew}
\end{figure}

\begin{figure}
\centering
\begin{tabular}{c}
\includegraphics[bb=112 397 458 614, clip=,width=0.90\hsize]{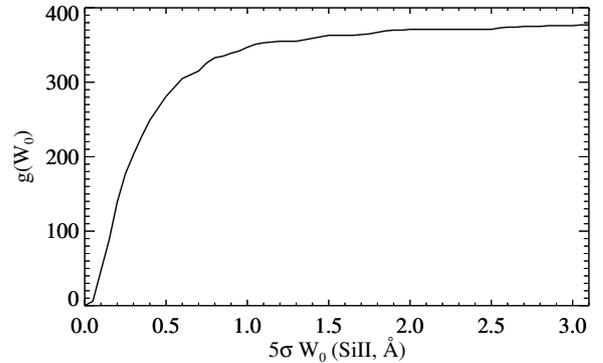}
\end{tabular}
\caption{The total redshift path density, $g(W_{0})$, as a function of the 5\,$\sigma$ Si\,{\sc ii} EW detection threshold.} 
 \label{gw}
\end{figure}

In Fig.\,\ref{si2ew}, we plot the Si\,{\sc ii}\,$\lambda$1526 rest equivalent width (EW) distribution for the eclipsing DLAs from our statistical sample. We choose Si\,{\sc ii}\,$\lambda$1526 transition because it is located in a region where SNR is usually high, and also the contamination by the telluric lines and sky emission is minimum. To evaluate the completeness of our Si\,{\sc ii} EW sample, we construct the \emph{redshift path density} function, $g(W)$, following the formalism used by \citet{1987ApJ...322..739L} and \citet{1992ApJS...80....1S}. This function gives the number of sight lines which a Si\,{\sc ii}\,$\lambda$1526 transition with rest-frame EW greater than or equal to $W$ could have been detected. Figure\,\ref{gw} shows the $g(W)$ function for our statistical sample. As shown in this figure, the Si\,{\sc ii} EW sample is not complete at low values (i.e. EW\,$<$\,1.0\,\textup{\AA}). For example, this sample is less than 80 per cent complete at EW\,$<$\,0.6\,\textup{\AA}. To correct for the lack of completeness at low EWs, we divide our Si\,{\sc ii} EW distribution by the normalized $g(W)$ function. The corrected distribution is shown as a black histogram in Fig.\,\ref{si2ew} while the red dashed histogram represents the distribution before correction.

We parameterize the corrected distribution by an exponential \citep[which was first used by][for Ly$\alpha$ forest absorption lines]{1980ApJS...42...41S},
\begin{equation}
N(W)dW = (\frac{N^{*}}{W^{*}})\,{\rm exp}\,(-\frac{W}{W^{*}})dW \,\, ,
\end{equation}
and a power law \citep[which was found by][to represent the EW distribution of Mg\,{\sc ii} absorption lines]{1987ApJS...64..667T},
\begin{equation}
N(W)dW = CW^{-\delta}\,dW \,\, ,
\end{equation}
where $N^{*}$ and $W^{*}$ (for the exponential) and $C$ and $\delta$ (for the power law) are 250.0, 0.52\,\textup{\AA}, 103.5, and 3.5, respectively.

As shown in Fig.\,\ref{si2ew}, the distribution deviates from both the exponential (green curve) and power law (blue curve) function at $\rm W^{\rm SiII}_{0}$\,$\lesssim$\,1.0\,\textup{\AA}. Since the distribution was already corrected for the effect of SNR (using the $g(W)$ function), one could attribute the break in the Si\,{\sc ii} EW distribution to the inability of our metal template technique (see section\,\ref{searchingmethod}) in finding DLAs with weak metal absorption lines. However, this scenario is ruled out as we also employed DLA template technique to search for those eclipsing DLAs with no or very weak metal absorption that could have been missed by our metal template technique. As discussed in section\,\ref{completeness}, the DLA template approach confirms that indeed only very few systems with log\,$N$(H\,{\sc i})\,$\ge$\,21.10 may be missed by the metal template technique. Therefore, the observed break in the Si\,{\sc ii} EW distribution function could imply that the deficit of systems with log\,$N$(H\,{\sc i})\,$\ge$\,21.10 and $\rm W^{\rm SiII}_{0}$\,$<$\,1.0\,\textup{\AA} is indeed real.

\subsection{Characterising the statistical sample}

Our statistical sample contains two kinds of eclipsing DLAs: those with a Ly$\alpha$ NEL in their DLA absorption troughs, and those with no such emission. When a Ly$\alpha$ NEL is seen in an eclipsing DLA trough, it implies that the transverse size of the cloud is larger than the BLR but at the same time smaller than the NLR and/or star-forming regions in the quasar host galaxy. If the DLA size were smaller than the BLR then the leaked emission from this region would elevate the flux level at the bottom of the DLA. In extreme cases where the DLA is sufficiently smaller than the BLR, the leaked emission from the BLR would fully fill the DLA trough, resulting in the formation of a ghostly DLA \citep{2017MNRAS.466L..58F}. On the other hand, if the covering factor of the DLA were 100 per cent then it would block all Ly$\alpha$ emission from the NLR and/or star-forming regions in the host galaxy, and consequently no Ly$\alpha$ NEL would be observed in the DLA trough.

Although a neighboring galaxy could also be a counterpart to the eclipsing DLAs with (at least weak) Ly$\alpha$ NEL, we favor the hypothesis that these DLAs are some clouds of neutral gas in the quasar host galaxy (see section\,\ref{sectcompositeintro}). Supporting this scenario, \citet{2009ApJ...693L..49H} suggest that the proximate DLA with a Ly$\alpha$ emission in its trough found towards the quasar SDSS\,J1240+1455 could arise from neutral hydrogen clouds in the host galaxy. Moreover, \citet{2017MNRAS.466L..58F} also discovered a compact ($\le$\,0.32\,pc) dense ($n_{\rm HI}$\,$\ge$\,1000\,cm$^{-3}$) cloud of neutral hydrogen gas very close to the central quasar ($\ge$\,39\,pc).

The clouds giving rise to the eclipsing DLAs with NEL could be located anywhere from the innermost region of the central quasar to the outermost region of the halo of the host galaxy. The origin of the gas could be infalling and/or outflowing. Here, we compare different properties of the eclipsing DLAs with and without NEL to investigate whether they are intrinsically different.

\subsubsection{Metals and dust}\label{sect_metal}

The rest EW of C\,{\sc ii}\,$\lambda$1334, Si\,{\sc ii}\,$\lambda$1526, Al\,{\sc ii}\,$\lambda$1670, and Fe\,{\sc ii}\,$\lambda$2344 measured for each eclipsing DLA in the full sample are listed in Table\,\ref{bigtable2}. In this table, we report the EWs determined from the Voigt profile fit to the absorption lines as well as the EWs measured directly from the observed spectra by integrating the pixel over the line profile. Since C\,{\sc ii} absorption line profile is usually blended with the C\,{\sc ii}$^{*}$\,$\lambda$\,1335 absorption, we only report in Table\,\ref{bigtable2}, the EWs directly measured from the normalized spectra and not from the Voigt profile fits.

\begin{figure}
\centering
\begin{tabular}{c}
\includegraphics[bb=126 398 458 649, clip=,width=0.90\hsize]{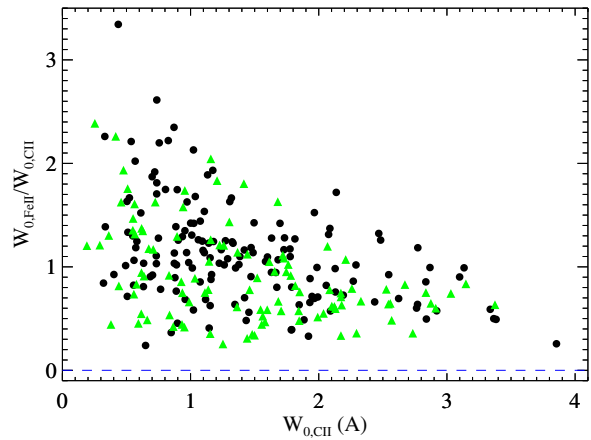}
\end{tabular}
\caption{Rest EW ratio W$_{0,\,\rm FeII}$/W$_{0,\,\rm CII}$  as a function of W$_{0,\,\rm CII}$. Eclipsing DLAs with (resp. without) narrow Ly$\alpha$ emission are shown as green filled triangles (resp. black filled circles).}
 \label{FeCvsC}
\end{figure}

In Fig.\,\ref{FeCvsC}, we plot the W$_{0,\,\rm FeII}$/W$_{0,\,\rm CII}$ ratios as a function of W$_{0,\,\rm CII}$ for the eclipsing DLAs with (green filled triangles) and without (black filled circles) narrow emission. Although the W$_{0,\,\rm FeII}$/W$_{0,\,\rm CII}$ ratios seem to decrease with increasing W$_{0,\,\rm CII}$, the ratios appear to be consistently smaller in DLAs with narrow emission. If we assume that the W$_{0,\,\rm FeII}$/W$_{0,\,\rm CII}$ ratio is an indication of the depletion in the gas \citep{2013A&A...558A.111F}, then the observed trend would imply that dust depletion is higher in eclipsing DLAs with narrow emission. We note that the same behavior is also seen when C\,{\sc ii} is replaced by Si\,{\sc ii}.

\begin{table}
\caption{First column: the J2000 coordinates of the quasars. Second column: quasar emission redshift. Third column: logarithm of H\,{\sc i} column density. Fourth column: absorption redshift measured from fitting low ionization metal absorption lines. Fifth column: redshift of the narrow Ly$\alpha$ emission. No redshift is reported when no emission is detected in the DLA trough. Sixth column: the integrated flux (IF) of the narrow Ly$\alpha$ emission. The unit is erg\,s$^{-1}$\,cm$^{-2}$. In cases where no Ly$\alpha$ emission is detected in the DLA trough we report an upper limit which is determined by integrating the pixel values in the DLA trough over a velocity extent of 1000\,km\,s$^{-1}$. Seventh column: DLAs from the statistical sample are labeled as 'S' while others are shown with 'N'. The full version of this table is available on-line. Only the first 15 entries are shown here.}
\centering 
 \setlength{\tabcolsep}{2.3pt}
\renewcommand{\arraystretch}{1.05}
\begin{tabular}{c c c c c c c} 
\hline\hline 
SDSS name & $z_{\rm QSO}$ & log\,$N$(H\,{\sc i}) & $z_{\rm abs}$ & $z_{Ly\alpha}$ & IF  &  sample\\ [0.5ex] 
\hline 
004309.69$-$011237.0 & 2.531 & 21.15 & 2.538 & ... & $\le$0.4 & S  \\
013654.00$-$005317.5 & 2.489 & 21.38 & 2.493 & ... & $\le$2.8 & S  \\
074359.68+394145.2 & 2.348 & 21.40 & 2.356 & ... & $\le$1.6 & S  \\
075057.94+403019.6 & 2.211 & 21.01 & 2.212 & ... & $\le$2.4 & N  \\
081224.37+442217.2 & 3.471 & 21.12 & 3.467 & 3.467 &  1.4 & S  \\
081343.14+481805.1 & 2.590 & 21.44 & 2.595 & ... & $\le$17.6 & S  \\
080956.02+502000.9 & 3.282 & 21.03 & 3.285 & 3.284 &  53.6 & N  \\
080045.73+341128.9 & 2.322 & 21.47 & 2.328 & 2.327 &  17.5 & S  \\
080052.00+315333.6 & 2.184 & 21.09 & 2.182 & ... & $\le$5.6 & N  \\
081626.02+351055.2 & 3.058 & 21.41 & 3.070 & 3.065 &  8.9 & S  \\
100112.40$-$014737.9 & 2.188 & 21.26 & 2.199 & 2.198 &  10.9 & S  \\
082741.29+404908.0 & 2.203 & 21.05 & 2.189 & ... & $\le$3.4 & N  \\
082755.89+400128.9 & 2.260 & 21.15 & 2.266 & ... & $\le$4.3 & S  \\
084448.59+043504.7 & 2.375 & 21.38 & 2.378 & ... & $\le$9.3 & S  \\
085600.89+030218.7 & 3.048 & 21.26 & 3.056 & 3.058 &  1.8 & S  \\
\hline 
\end{tabular}
\label{bigtable1} 
\end{table}

\begin{table*}
\caption{Rest EWs of C\,{\sc ii}$\lambda$1334, Si\,{\sc ii}$\lambda$1526, Al\,{\sc ii}$\lambda$1670 and Fe\,{\sc ii}$\lambda$2344 transitions. For each species, the first column (labeled as 'F') is the EW measured directly by integrating the pixel values over the line profile; the second column (labeled as 'Fit') is the EW determined from a fit to the absorption line; and the third column (labeled as 'Err') gives the uncertainty on the measured EW. Since C\,{\sc ii} absorption is typically blended with C\,{\sc ii}$^{*}$, we only report the EW of the whole C\,{\sc ii}\,+\,C\,{\sc ii}$^{*}$ absorption feature determined from direct integration of the pixel values. Moreover, when the spectral region of a transition is too noisy, no EW is reported for that species. The full version of this table is available on-line. Only the first 10 entries are shown here.} 
\centering 
 \setlength{\tabcolsep}{3.0pt}
\renewcommand{\arraystretch}{1.05}
\begin{tabular}{c c c c c c c c c c c c c c} 
\hline\hline 

SDSS name & $z_{\rm abs}$ &   &  C\,{\sc ii}\,$-$\,1334  &   &   & Si\,{\sc ii}\,$-$\,1526 &  &   & Al\,{\sc ii}\,$-$\,1670  &   &    & Fe\,{\sc ii}\,$-$\,2344 &  \\
          &               & F & Fit & Err &  F & Fit & Err & F & Fit & Err & F & Fit & Err \\
\hline
004309.69$-$011237.0 & 2.538 & 1.488 & ... & 0.396 & 1.606 & 1.614 & 0.252 & 2.703 & 2.823 & 0.802 & ... & ... & ...  \\
013654.00$-$005317.5 & 2.493 & 2.086 & ... & 0.174 & 1.797 & 1.927 & 0.088 & 2.146 & 2.081 & 0.173 & 2.715 & 2.862 & 0.205  \\
074359.68+394145.2 & 2.356 & 0.703 & ... & 0.173 & 0.384 & 0.384 & 0.100 & 0.621 & 0.589 & 0.120 & 0.685 & 0.647 & 0.142  \\
075057.94+403019.6 & 2.212 & 2.582 & ... & 0.151 & 1.539 & 1.647 & 0.056 & 1.455 & 1.470 & 0.051 & 1.638 & 1.691 & 0.072  \\
081224.37+442217.2 & 3.467 & 0.674 & ... & 0.194 & 0.630 & 0.632 & 0.052 & 0.878 & 0.876 & 0.070 & ... & ... & ...  \\
081343.14+481805.1 & 2.595 & 0.755 & ... & 0.173 & 0.948 & 1.055 & 0.163 & 1.120 & 1.187 & 0.304 & 1.762 & 1.660 & 1.341  \\
080956.02+502000.9 & 3.285 & 0.413 & ... & 0.053 & 0.639 & 0.606 & 0.035 & 0.372 & 0.303 & 0.033 & 0.512 & 0.515 & 0.170  \\
080045.73+341128.9 & 2.328 & 0.988 & ... & 0.170 & 0.724 & 0.782 & 0.083 & 0.734 & 0.751 & 0.113 & 0.628 & 0.653 & 0.171  \\
080052.00+315333.6 & 2.182 & 1.542 & ... & 0.634 & 0.967 & 0.972 & 0.216 & 0.508 & 0.629 & 0.220 & ... & ... & ...  \\
\hline 
\end{tabular}
\label{bigtable2} 
\end{table*}

\subsubsection{Kinematics}

For each quasar, we checked the z_PCA (i.e. the redshift automatically determined using the PCA method) and z_vis (i.e. the redshift from the visual inspection) from the \citet{2017A&A...597A..79P} catalog and found that some of the reported redshifts are not reliable enough. For example, in some cases the redshifts in the catalog do not correspond to any quasar intrinsic emission lines detected in the spectra. We, therefore, re-measure the quasars emission redshift by fitting Gaussian function on the available intrinsic emission lines, such as C\,{\sc iii},  Mg\,{\sc ii}, and He\,{\sc ii}. We avoid using C\,{\sc iv} emission line as it is not a good quasar redshift indicator \citep{1992ApJS...79....1T}. 

\begin{figure}
\centering
\begin{tabular}{c}
\includegraphics[bb=120 398 459 649, clip=,width=0.90\hsize]{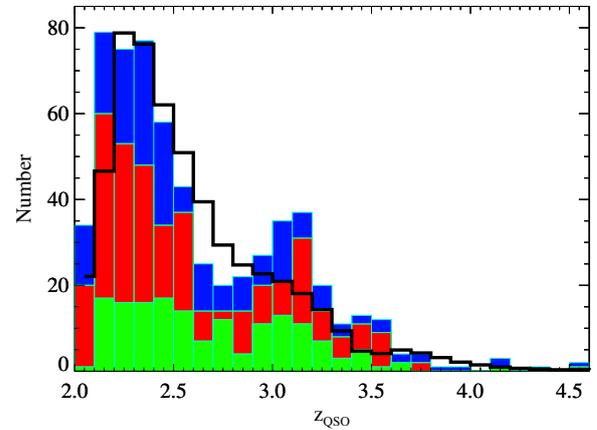}
\end{tabular}
\caption{Distribution of quasar redshifts for the full (blue histogram) and statistical (red histogram) samples. The green histogram shows the distribution for the DLAs with NELs in the statistical sample. For comparison, the distribution for all DR12 quasars with $z_{\rm QSO}$\,$>$\,2.0 is also shown as black histogram. Note that the black histogram is normalized to the maximum value of the blue histogram.}
 \label{zemdist}
\end{figure}

Figure\,\ref{zemdist} shows the quasar emission redshift distributions for the full and statistical samples, along with the redshift distribution for all DR12 quasars with $z_{\rm QSO}$\,$>$\,2.0. The striking feature in this figure is the excess of quasars at $z_{\rm QSO}$\,=\,3.0--3.2 in the full and statistical sample. This excess which was also reported by \citet{2013A&A...558A.111F}, is not seen in the DR12 quasar redshift distribution (see black histogram in Fig.\,\ref{zemdist}). Moreover, the distribution for the quasars (from the statistical sample) that have DLAs with NEL (i.e. the green histogram) is mostly flat, implying that there is no preferred redshift for the occurrence of a NEL in an eclipsing DLA trough.

\begin{figure}
\centering
\begin{tabular}{c}
\includegraphics[bb=120 398 461 649, clip=,width=0.90\hsize]{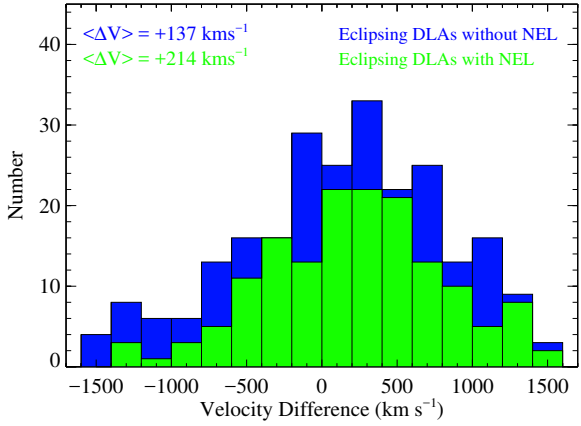}
\end{tabular}
\caption{Velocity difference distribution for $z_{\rm DLA}$ relative to $z_{\rm QSO}$. $z_{\rm DLA}$ is the absorption redshift measured from the fit to low ionization absorption lines. Eclipsing DLAs that exhibit narrow Ly$\alpha$ emission are shown in green while those without such emission are in blue. There appears to be a preference for positive velocity offset in DLAs revealing narrow emission. Although positive velocity offset could imply infalling gas, uncertain quasar emission redshift determination could also be an explanation.}
 \label{dV_DLA_QSO}
\end{figure}

Figure\,\ref{dV_DLA_QSO} presents the velocity offset between the eclipsing DLAs and the quasars from our statistical sample. Here, positive velocity offset means that the DLA is redshifted with respect to the quasar. As shown in this figure, for both eclipsing DLAs with and without a NEL, there appears to be more DLAs with positive velocity offset. Indeed, $\sim$\,70 (resp. $\sim$\,60) per cent of the eclipsing DLAs with (resp. without) a NEL exhibit positive velocity offset with respect to their quasars. The mean velocity offset for the DLAs with narrow emission is a factor of $\sim$\,1.6 larger than that of the other population. This could indicate that eclipsing DLAs with narrow emission may be part of some infalling gas.
However, since redshifts determined from the quasar broad emission lines usually underestimate the quasar systemic redshift \citep{2010MNRAS.405.2302H}, the observed positive velocity offset could also be a consequence of uncertain quasar redshift estimation.  Follow-up infrared observations of these quasars would confirm whether the gas is infalling by precisely measuring the quasar emission redshift using some narrow emission lines such as [O\,{\sc ii}] and [O\,{\sc iii}].

\begin{figure}
\centering
  \setlength{\unitlength}{1cm}
  \begin{picture}(20, 5.3)
     \put(0,-0.4)       {\includegraphics[bb = 95 398 458 720, clip=, width=0.8\hsize]{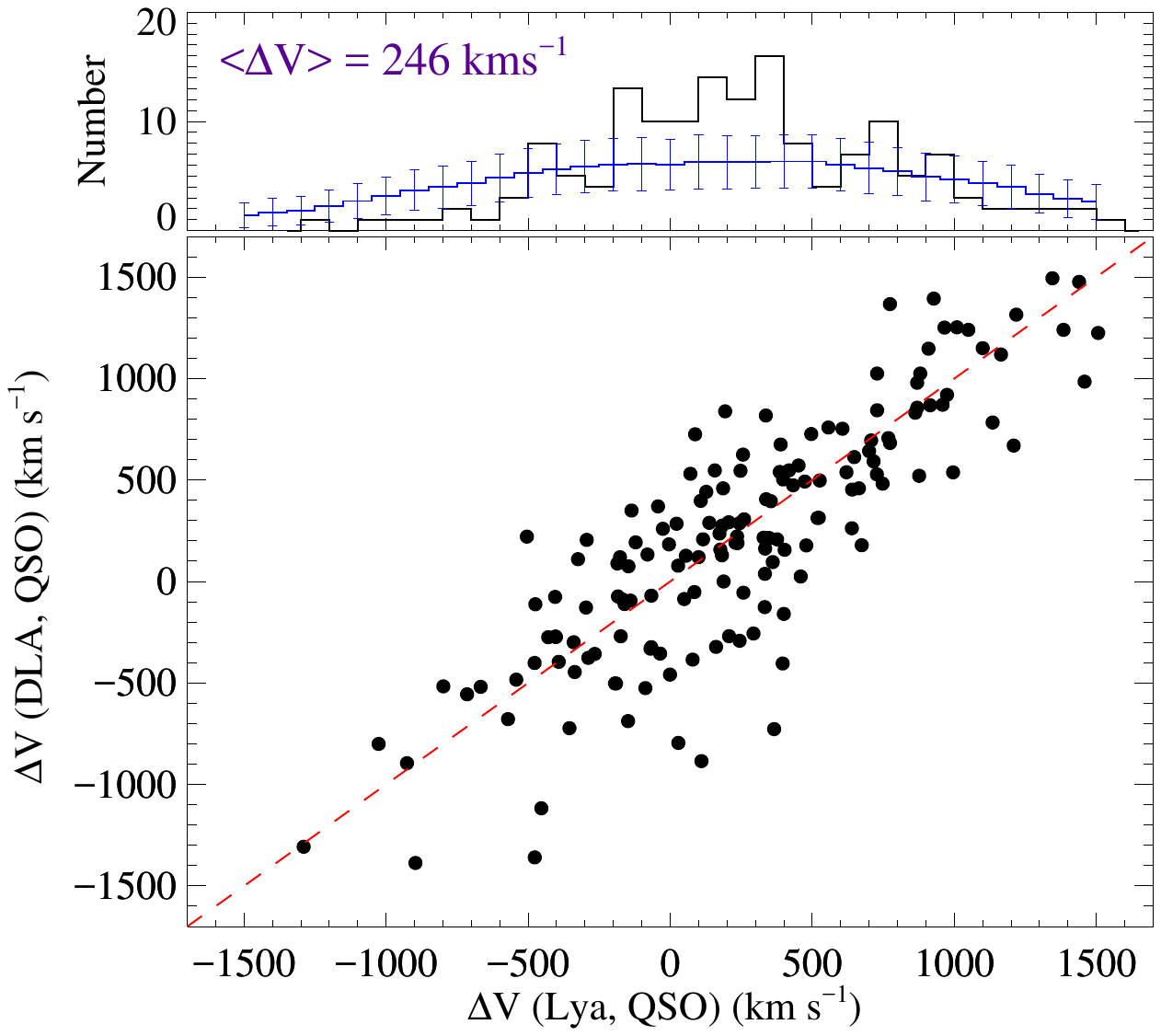}}
     \put(6.524,4.335)  {\includegraphics[bb = 146 610 501 719, clip=, width=0.555\hsize, angle=-90]{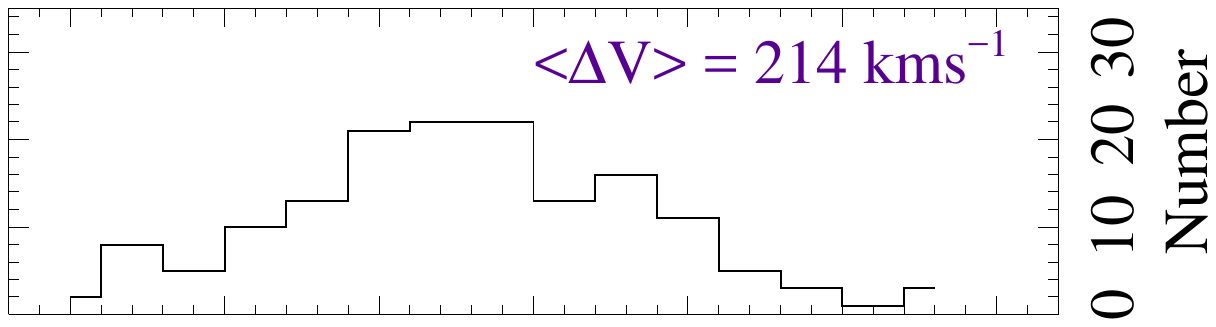}}
      \end{picture}
   \caption{Velocity offset between eclipsing DLAs and the quasars as a function of the velocity difference between the narrow Ly$\alpha$ emission and the quasars. The dashed red line show the regions where $\Delta$V(Ly$\alpha$, QSO)\,=\,$\Delta$V(DLA, QSO). The panels on the right and top show their corresponding distributions. The blue histogram in the upper panel shows the results of Monte Carlo simulations for the NEL-QSO velocity offset distribution. The simulation shows that the NEL may not be randomly distributed in the eclipsing DLAs trough.}
  \label{dvsimul}
 \end{figure}

We show in Fig.\,\ref{dvsimul}, the velocity offset of the Ly$\alpha$ NEL relative to the quasar as a function of the DLA-QSO velocity offset. Since  the Ly$\alpha$ narrow emission line does not always exhibit a symmetric one-component feature (see Fig.\,\ref{example_plots}), we therefore estimate its redshift by weighting the wavelength in each pixel by the flux of that pixel.
As shown in Fig.\,\ref{dvsimul}, there appears to be a correlation between the two velocity offsets, with the correlation coefficient of $\sim$\,0.84. This correlation could indicate a connection between the NEL and the eclipsing DLA.
The dispersion seen in the $\Delta$V(Ly$\alpha$, QSO) and $\Delta$V(DLA, QSO) values could be due to the uncertainty in the quasar emission redshift. The NEL-QSO (resp. DLA-QSO) velocity offset distribution is shown in the small upper (resp. right hand side) panel in Fig.\,\ref{dvsimul}. Interestingly, the mean velocity offset of the two distribution is almost the same.

\begin{figure}
\centering
\begin{tabular}{c}
\includegraphics[bb=120 398 461 650, clip=,width=0.90\hsize]{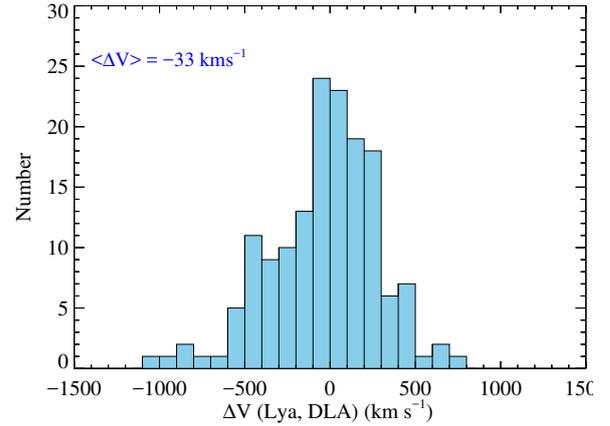}
\end{tabular}
\caption{Velocity offset distribution for $z_{\rm Ly}$$_{\alpha}$ relative to $z_{\rm DLA}$ for the statistical sample of eclipsing DLAs with narrow Ly$\alpha$ emission. }
 \label{dV_Lya_DLA}
\end{figure}

We perform a Monte Carlo simulation to check whether the random placement of the NEL inside the DLA trough could reproduce the observed distribution. The result of 2000 Monte Carlo simulations is shown as a blue histogram in the upper panel of Fig.\,\ref{dvsimul}. The error on the distribution in each velocity bin is the standard deviation measured from the simulations. The blue histogram shows a nearly flat distribution from $-$500 to 800\,km\,s$^{-1}$ with no peak around a specific velocity while the observed histogram seems to peak around 300\,km\,s$^{-1}$. We also plot in Fig.\,\ref{dV_Lya_DLA} the NEL-DLA velocity offset distribution. The distribution peaks around zero velocity ($\Delta$V\,$\sim$\,$-$33\,km\,s$^{-1}$). This, along with the NEL-QSO distribution shown in the upper panel of Fig.\,\ref{dvsimul}, could be an indication that the NEL and their corresponding eclipsing DLA both originate in the quasar host galaxy.

\subsection{Ly$\alpha$ narrow emission luminosities}

\begin{figure}
\centering
\begin{tabular}{c}
\includegraphics[bb=115 398 464 649, clip=,width=0.90\hsize]{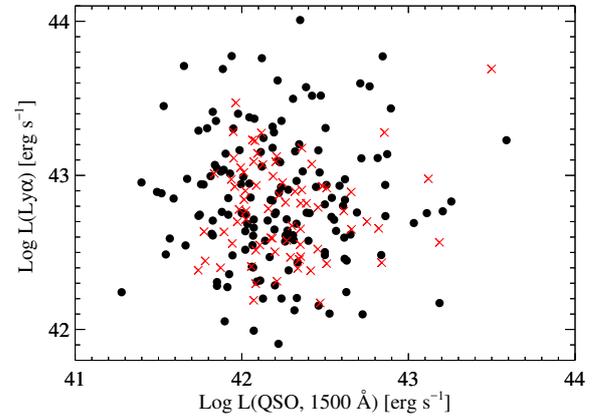}
\end{tabular}
\caption{Luminosities of the Ly$\alpha$ NEL as a function of the quasar luminosity at 1500\,\textup{\AA}. Filled black circles mark the systems from the statistical sample, and the red crosses represent systems with log\,$N$(H\,{\sc i})\,$<$\,21.10.}
\label{Lqso_nlr}
\end{figure}

\begin{figure}
\centering
\begin{tabular}{c}
\includegraphics[bb=112 398 458 649, clip=,width=0.90\hsize]{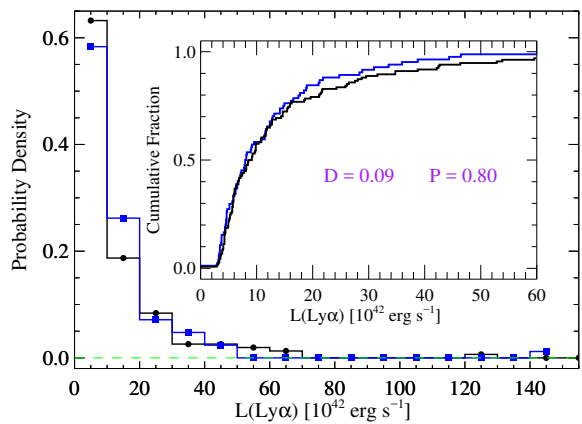}
\end{tabular}
\caption{Distribution of Ly$\alpha$ emission luminosities for our statistical sample (black histogram) and LAEs \citep[blue histogram; ][]{2008ApJS..176..301O}. The inset plot shows the KS test of the two distributions with the same color scheme. The KS test plot is truncated at 60\,$\times$\,10$^{42}$\,erg\,s$^{-1}$.}
 \label{nlr_distribution}
\end{figure}

We show, in Fig.\,\ref{Lqso_nlr}, the luminosity of the Ly$\alpha$ NEL as a function of the quasar luminosity at 1500\,\textup{\AA} in the rest frame. As shown in this figure, no correlation is found between the two luminosities. The lack of a correlation could indicate that each eclipsing DLA covers random fraction of the Ly$\alpha$ emitting region.  We also compare, in Fig.\,\ref{nlr_distribution}, the luminosity distribution of the Ly$\alpha$ NEL detected in the troughs of the eclipsing DLAs from our statistical sample (black histogram) to that of Ly$\alpha$ emitters (LAEs; blue histogram) from \citet{2008ApJS..176..301O}. LAEs are low-mass high-redshift galaxies that are believed to be actively forming stars \citep{2009NewAR..53...37N}. The Ly$\alpha$ emission from LAEs is primarily due to star formation although some AGN activity has also been reported in some LAEs, especially in those with high luminosities (Log\,L(Ly$\alpha$)\,>\,43.5\,erg\,s$^{-1}$, see \citet{2008ApJS..176..301O}; \citet{2013ApJ...771...89O}). From the cumulative distribution in Fig.\,\ref{nlr_distribution}, one can see that the two populations have similar distributions up to about 10\,$\times$\,10$^{42}$\,erg\,s$^{-1}$. The fraction of objects with higher luminosity is larger in our sample. This implies that on the basis of these distributions alone, we cannot distinguish between our Ly$\alpha$ NELs and Ly$\alpha$ emitters except for the strongest systems.

For the weak systems, we do not find a difference between Ly$\alpha$ emission from star formation and from extended emission around the quasar. Nonetheless, we favor the second hypothesis. Here, the eclipsing DLA is larger than the quasar central NLR but is small enough so that emission from fluorescence in the outskirts of the host galaxy can be seen \citep[see][]{2015MNRAS.454..876F,2016MNRAS.461.1816F}. For the largest luminosities however it is clear that the eclipsing DLA covers only part of the NLR of the quasar.  In this case, the DLA is probably smaller in size and closer to the quasar compared to the DLAs that reveal weaker and/or no emission. If the eclipsing DLAs with stronger Ly$\alpha$ emission are indeed closer to the quasar and smaller in size, we then expect their absorption line properties to be different from those of the eclipsing DLAs that reveal no or weak emission. Therefore, in the next section, we construct composite spectra of the eclipsing DLAs with no, weak and strong Ly$\alpha$ NEL in their troughs, in order to probe the validity of this scenario.

\subsection{Composite spectra} \label{sectcompositeintro}

In this section, we create a number of stacked spectra using the quasar spectra from
our statistical sample in order to statistically look for differences between the eclipsing DLAs that reveal no, weak, and strong Ly$\alpha$ NEL in their DLA troughs.
We first compare composite spectra for eclipsing DLAs with and without narrow Ly$\alpha$ emission detected (section\,\ref{sectcomposite1}), and then we further subdivide the emission sample and investigate composite spectra for eclipsing DLAs with weak and strong narrow emission (section\,\ref{sectcomposite2}).
Before generating composite spectra, we shift all spectra to the rest frame of the DLA absorbers. The spectra are then normalized by  iteratively fitting the quasars continua using the Savitzky-Golay filtering \citep{1964AnaCh..36.1627S} and removing deviant pixels.
Composites are finally generated by median combining the rest-frame normalized spectra \citep{2010MNRAS.406.1435E}. Compared with a mean combination, the median has the advantage of eliminating more robustly the spikes, cosmic ray hits and also absorbers with very strong absorption lines. However, we also generated mean composites and found that the basic results are unchanged.

\subsubsection{Composites of eclipsing DLAs with and without NEL} \label{sectcomposite1}

\begin{figure*}
\centering
\begin{tabular}{cc}
\includegraphics[bb=51 352 554 656, clip=,width=0.90\hsize]{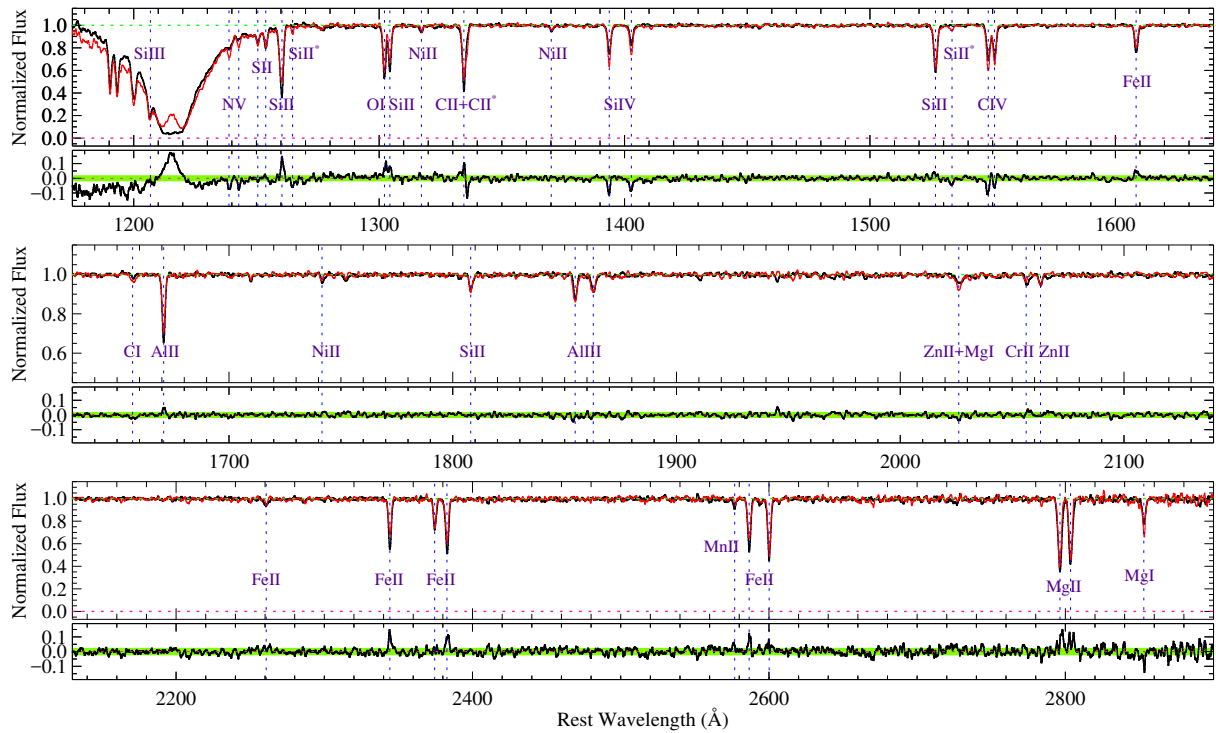}
\end{tabular}
\caption{Median-combined stacked spectra (in the absorber rest frame) of the 155 eclipsing DLAs with narrow emission (red spectrum) and the 244 eclipsing DLAs with no such emission (black spectrum) from our statistical sample. Below each main panel, we also plot the difference spectrum which is the difference of the two composites. Here the black spectrum is subtracted from the red one. The green strip shows the 1\,$\sigma$ error.}
 \label{stackedfull}
\end{figure*}

\begin{figure*}
\centering
\begin{tabular}{c}
\includegraphics[bb=73 367 524 645, clip=,width=0.90\hsize]{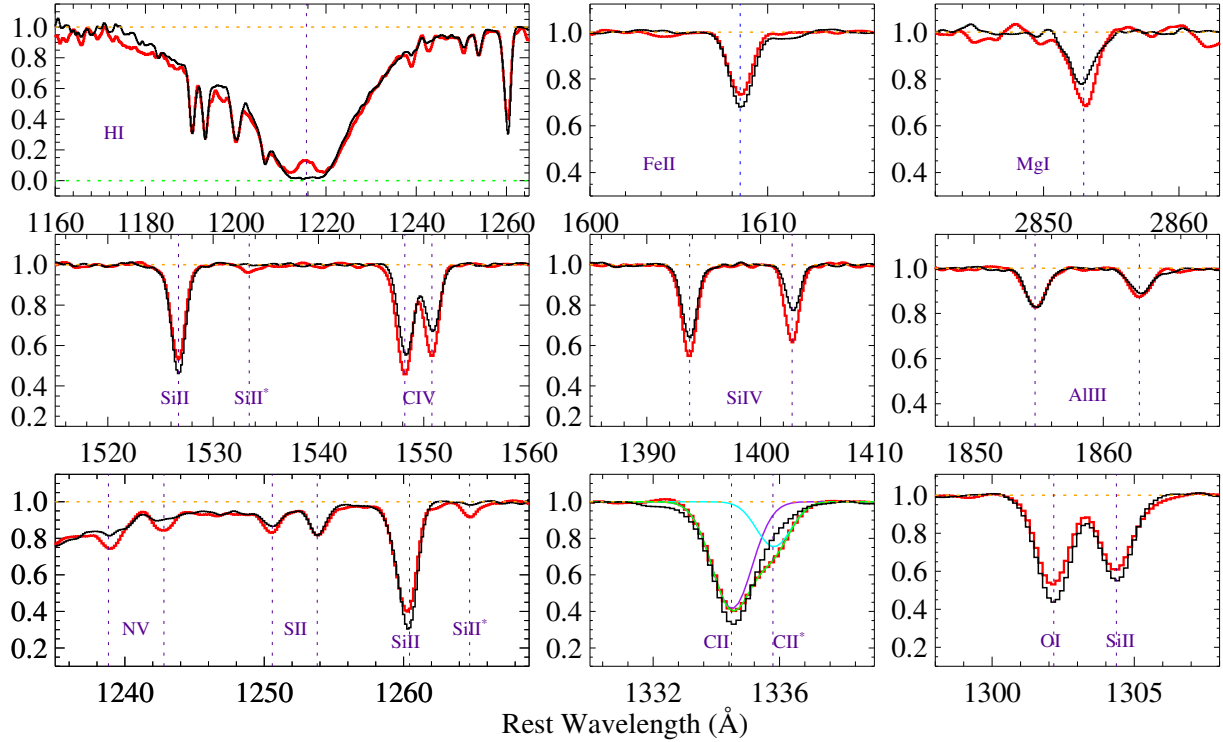}
\end{tabular}
\caption{Some important transitions detected in the composite spectra of eclipsing DLAs with (red curves) and without (black curves) narrow Ly$\alpha$ emission are shown in greater details. In the C\,{\sc ii} velocity panel, we decompose the absorption feature by conducting a double-Gaussian fit. Here, the purple (resp. cyan) curve represents the C\,{\sc ii} (resp. C\,{\sc ii}$^{*}$) component. The final fit is overplotted as a green curve.}
 \label{zoomedplots1}
\end{figure*}

In this section, we investigate whether the eclipsing DLAs with Ly$\alpha$ NEL are denser and physically closer to their background quasars. To this end, we create composite spectra for the eclipsing DLAs with (containing 155 spectra) and without (containing 244 spectra) a NEL in their DLA troughs, and compare the strength of the absorption lines in the two composites. Hereafter, eclipsing DLAs with (resp. without) a NEL are also called EPDLAs (resp. NEPDLAs). Since the two subsamples have different sizes, we check whether this has any significant effect on the results. To check this, we randomly chose 155 spectra (out of 244) from the NEPDLA sample and stacked them. We repeated this process 500 times and found that the EWs of the absorption lines do not change by more than 5 per cent.
We, therefore, use the full sample composites in this work. The two composite spectra are shown in Fig.\,\ref{stackedfull}, with the black and red spectra representing the NEPDLA and EPDLA composites, respectively. In this figure, the small panel below each main panel shows the difference spectrum, which is the difference of the two composites. Here, the black spectrum is subtracted from the red one. When an absorption (resp. emission) like feature occurs in the difference spectrum at a certain wavelength, it implies that the absorption line corresponding to that wavelength is stronger (resp. weaker) in the EPDLA (resp. NEPDLA) composite. For better visualization, we magnified in Fig.\,\ref{zoomedplots1}, the spectral regions corresponding to some important transitions (e.g. C\,{\sc ii}, Mg\,{\sc i}, ...). Note that the median log\,$N$(H\,{\sc i}) of the eclipsing DLA absorbers in the NEPDLA and EPDLA samples is 21.30 and 21.37, respectively.

\begin{figure}
\centering
\begin{tabular}{ll}
\includegraphics[bb=112 398 465 595, clip=,width=0.90\hsize]{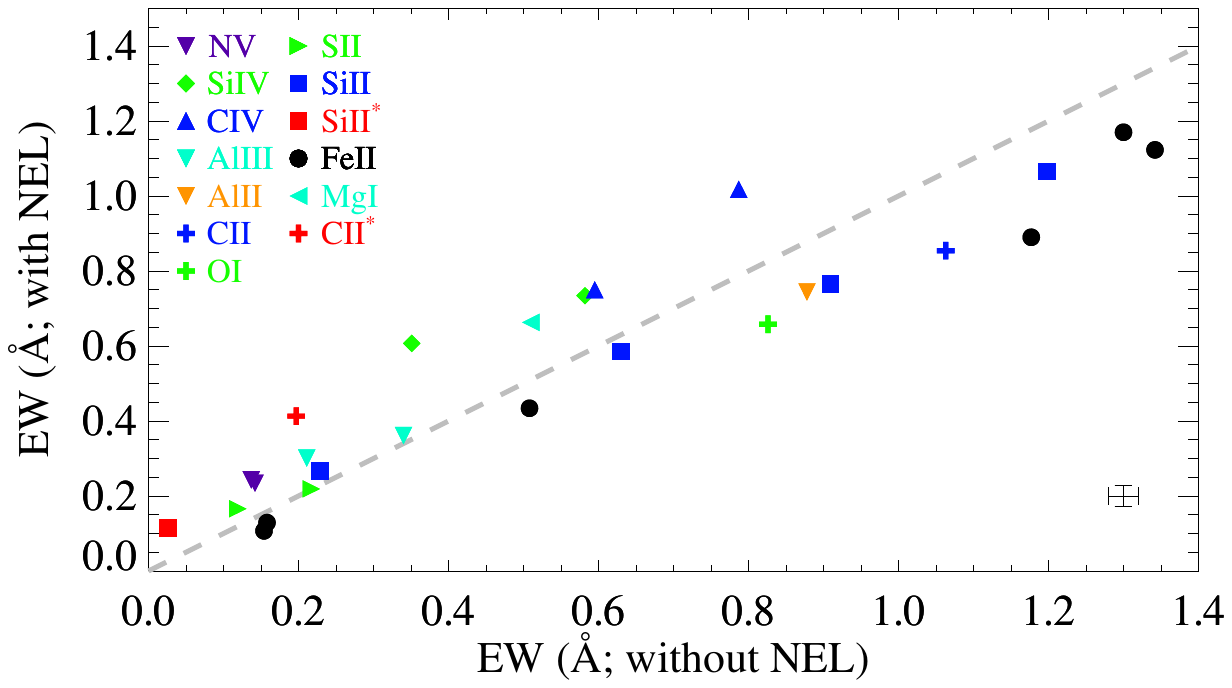} \\
\includegraphics[bb=112 398 465 595, clip=,width=0.90\hsize]{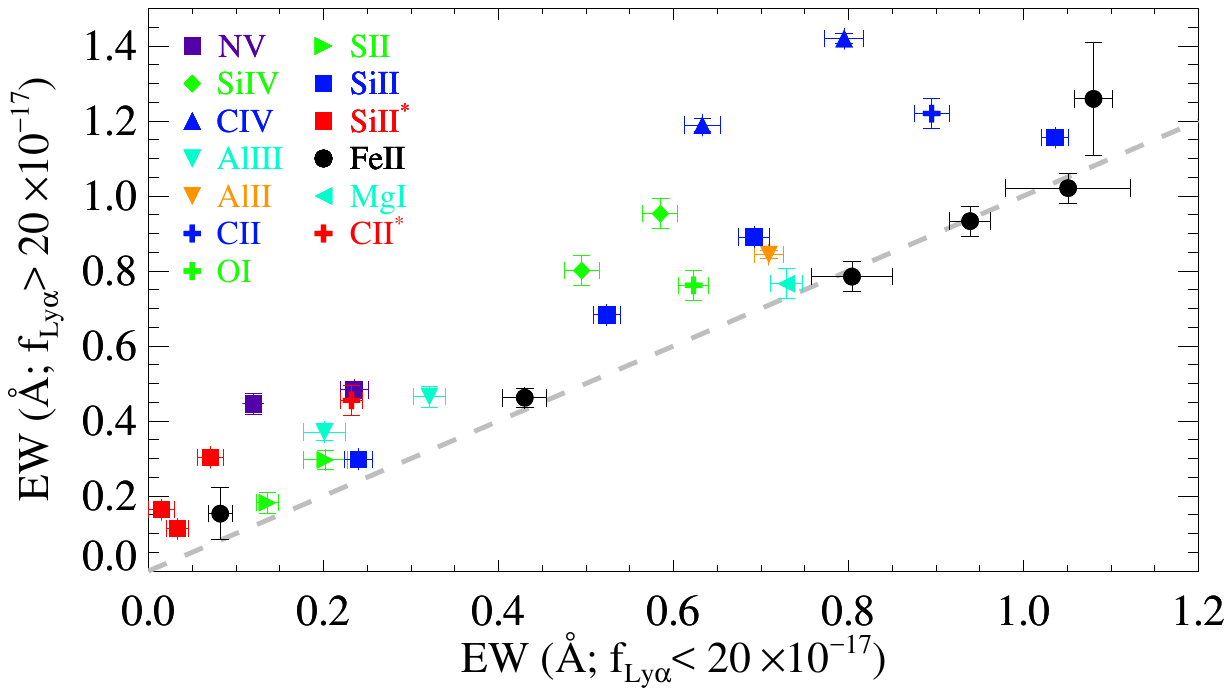}
\end{tabular}
\caption{Upper panel: EWs of the absorption lines detected in the EPDLA composite as a function of those detected in the NEPDLA composite. The error bar on the lower right side shows the uncertainty in the EWs of the data points. Lower panel: EWs of the absorption lines detected in the composite of DLAs with strong narrow emission as a function of those detected in the composite of DLAs with weak narrow emission (see the text for the definition of the weak and strong narrow emission). The dashed grey curves in the two panels mark the regions where the two EWs are the same.}
 \label{EWvsEW}
\end{figure}

We determined the EWs of the absorption lines using single Gaussian fits. In cases where the absorption lines were blended (i.e. the C\,{\sc iv} doublet, C\,{\sc ii}, C\,{\sc ii}$^{*}$, O\,{\sc i} and Si\,{\sc ii} absorption lines) a two-component Gaussian fits was used \citep{2013MNRAS.435.1727F}. The rest-frame EWs of the detected transitions are listed in Table\,\ref{ewtable} (columns\,3\,\&\,4). In the upper panel of Fig.\,\ref{EWvsEW}, the EWs of the absorption lines detected in the EPDLA composite are plotted as a function of those detected in the NEPDLA composite, and the dashed grey curve marks the regions where the two EWs are equal. As shown in this figure, the EWs of the low ionization species (except for Mg\,{\sc i}) in the two composites are consistent with each other when the EWs are less than 0.7\,\textup{\AA}. On the other hand, all low ionization species with EWs larger than 0.7\,\textup{\AA} have stronger absorption in the NEPDLA composite. Since absorption lines with EWs\,$>$\,0.7 all fall on the saturated portion of the curve of growth, higher EWs of these transitions in the NEPDLA composite could be due to the higher Doppler-$b$ parameter.

\begin{figure}
\centering
\begin{tabular}{c}
\includegraphics[bb=155 398 413 685, clip=,width=0.90\hsize]{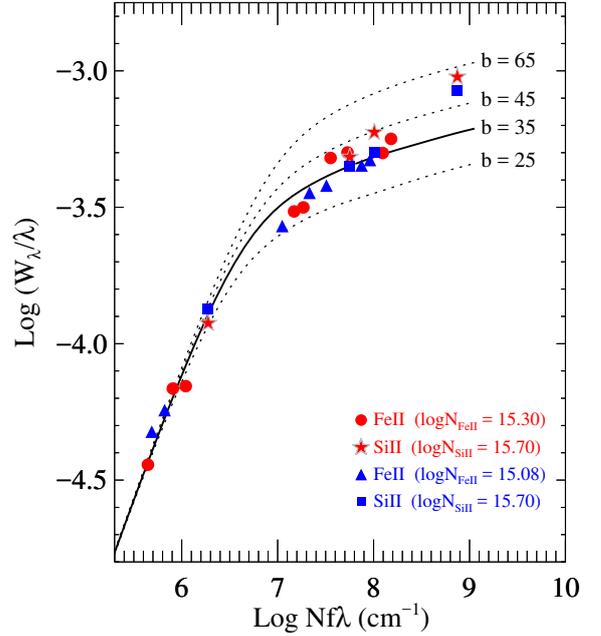}
\end{tabular}
\caption{Empirical curve of growth constructed using the Fe\,{\sc ii} absorption lines for the NEPDLA (red symbols) and EPDLA (blue symbols) composites. Column densities are from weak transitions and equivalent widths are from Table\,\ref{ewtable}.}
 \label{cogfe2}
\end{figure}

Although the low resolution of the SDSS spectra does not allow for a robust determination of the abundances, we still use the \citet{2008ApJ...685..344P} calibration (i.e. log\,$Z/Z_{\odot}$\,=\,$-$0.92\,+\,1.42\,log($W$) where $W$ is the EWs of the Si\,{\sc ii}\,$\lambda$1526 in \textup{\AA}) to have a rough estimate of the metallicities. From this calibration we get $-$0.98 and $-$1.08 as the metallicities of the NEPDLA and EPDLA samples, respectively. The uncertainty in the \citet{2008ApJ...685..344P} calibration is 20 per cent. Moreover, \citet{1986ApJ...301..355J} have demonstrated that a curve of growth can be constructed for a composite spectrum to derive reliable column densities, provided that some transitions of the species of interest fall on the linear part of the curve of growth. We, therefore, construct an empirical curve of growth using the Fe\,{\sc ii} absorption lines from the NEPDLA and EPDLA composite spectra (Fig.\,\ref{cogfe2}). The data from the Si\,{\sc ii} are also shown in Fig.\,\ref{cogfe2}. Interestingly the silicon abundances derived from the curve of growth analysis for the NEPDLA ([Si/H]\,=\,$-$1.16\,$\pm$\,0.20) and EPDLA ([Si/H]\,=\,$-$1.21\,$\pm$\,0.20) composites are consistent with those obtained from the calibration of \citet{2008ApJ...685..344P}.

We detect absorption from the excited states of Si\,{\sc ii} and C\,{\sc ii} in both composites. Although Si\,{\sc ii}$^{*}$\,$\lambda$1533 is only detected in the EPDLA composite, the Si\,{\sc ii}$^{*}$\,$\lambda$1264 seems to be weakly detected in the other composite. The Si\,{\sc ii}$^{*}$/Si\,{\sc ii} ratio from the EPDLA composite is a factor of $\sim$\,5.0 higher than in the other composite (see Table\,\ref{ewtable}). This may imply that the gas in the EPDLA systems is denser and/or exposed to a stronger radiation field. Since there are indications that the gas giving rise to EPDLAs are of higher densities (see below), collisional excitation and UV pumping could both be important in populating the fine structure states in EPDLA clouds.

The C\,{\sc ii}$^{*}$\,$\lambda$\,1335 absorption, on the other hand, is blended with the C\,{\sc ii}\,$\lambda$1334 absorption. We, therefore, performed a two-component Gaussian fit on the C\,{\sc ii} and C\,{\sc ii}$^{*}$ absorption feature in order to decompose the two absorption lines. The resultant fit is shown as a green curve in Fig.\,\ref{zoomedplots1}, with the purple and cyan curves representing the C\,{\sc ii} and C\,{\sc ii}$^{*}$ absorption, respectively. The equivalent widths measured from the fit are listed in Table\,\ref{ewtable}. From Table\,\ref{ewtable}, one can see that the C\,{\sc ii}$^{*}$/C\,{\sc ii} ratio is also higher (by a factor of $\sim$\,2.6) in the EPDLA composite, indicating higher density of the gas and/or close vicinity to an intense UV radiation source \citep{2002MNRAS.329..135S,2005MNRAS.362..549S,2008ApJ...681..881W}.

A striking feature in Fig.\,\ref{EWvsEW} is the EW of Mg\,{\sc i}, which is larger by $\sim$\,20 per cent in EPDLA composite. This is contrary to what is seen for other low ionization species. Since the two composites have almost similar median H\,{\sc i} column density, stronger Mg\,{\sc i} absorption could also indicate higher gas density.

Contrary to the low-ionization species, absorption from the high ionization species (i.e. N\,{\sc v}, Si\,{\sc iv} and C\,{\sc iv}) are all systematically stronger in the EPDLA composite (see Fig.\,\ref{zoomedplots1}). The EWs of Si\,{\sc iv}\,$\lambda$1393 and C\,{\sc iv}\,$\lambda$1548 are both $\sim$\,30 per cent larger in the EPDLA composite. The N\,{\sc v} absorption exhibits the highest difference between the two composites as it is $\sim$\,65 per cent larger in the EPDLA composite. Stronger N\,{\sc v} absorption could be due to higher metallicity \citep{2010MNRAS.406.1435E} and/or enhanced ionization \citep{2009A&A...503..731F}. Since the two composites seem to have similar metallicities, the stronger N\,{\sc v} absorption in the EPDLA composite could imply that the external layer of the cloud is more ionized. The stronger Si\,{\sc iv} and C\,{\sc iv} in the EPDLAs could also be attributed to the enhanced ionization. Moreover, although the EW of Al\,{\sc iii}\,$\lambda$1854 in the two composites are consistent with each other, the EW of Al\,{\sc ii} is smaller by $\sim$\,20 per cent in the EPDLA composite, indicating larger Al\,{\sc iii}/Al\,{\sc ii} ratio. Taken at face value, higher Al\,{\sc iii}/Al\,{\sc ii} ratio in the EPDLAs could indicate proximity to an ionizing source.

Stronger absorption from the high ionization species along with the higher Si\,{\sc ii}$^{*}$/Si\,{\sc ii} and C\,{\sc ii}$^{*}$/C\,{\sc ii} ratios in the EPDLA composite, all seem to be consistent with a scenario in which the DLA absorbers with narrow Ly$\alpha$ emission in their troughs are denser and closer to the quasar and perhaps located in the quasar host galaxy. The DLA absorbers with no such emission could be located farther away in a neighboring galaxy. A neighboring galaxy could also be a counterpart to the DLAs with weak Ly$\alpha$ NEL. In this case, the galaxy fully covers the NLR of the quasar but still allows the extended emission from the quasar host galaxy to pass them by unobscured. On the other hand, the DLAs that reveal strong Ly$\alpha$ NEL are highly likely located in the quasar host galaxy as they appear to be partially covering the NLR of the background quasar (see next section).

\subsubsection{Subsamples defined by Ly$\alpha$ NEL strength}  \label{sectcomposite2}

In this section, we would like to test the hypothesis that eclipsing DLAs exhibiting stronger narrow Ly$\alpha$ emission in their troughs are also denser and closer to the AGN (compared to the eclipsing DLAs with weak Ly$\alpha$ emission), and that when the emission is stronger, we are actually seeing part of the NLR of the quasar. To test this hypothesis, we divide the EPDLA sample into two subsamples based on the strength of the narrow Ly$\alpha$ emission. We then create composite spectra for the two subsamples and compare the properties of their absorption lines.
We select the integrated flux, IF\,=\,20\,$\times$\,10$^{-17}$\,erg\,s$^{-1}$\,cm$^{-2}$, to distinguish between the \emph{weak} and \emph{strong} emission.
We note that the basic conclusions stay the same if the cut is made at other IF. It is just that we chose
this value of IF because the most obvious difference in the absorption line properties of the composite spectra of the two subsamples are seen when the cut is made at this IF. The weak subsample (i.e. the subsample with IF\,$<$\,20\,$\times$\,10$^{-17}$\,erg\,s$^{-1}$\,cm$^{-2}$) contains 100 quasars while the strong subsample (i.e. the subsample with IF\,$\ge$\,20\,$\times$\,10$^{-17}$\,erg\,s$^{-1}$\,cm$^{-2}$) comprises 55 quasars. Following the same approach as in the previous section, we checked that different sample size has no significant effect on the results. The median H\,{\sc i} column density of the weak and strong subsamples is 21.35 and 21.49, respectively.

\begin{figure*}
\centering
\begin{tabular}{c}
\includegraphics[bb=73 367 523 645, clip=,width=0.90\hsize]{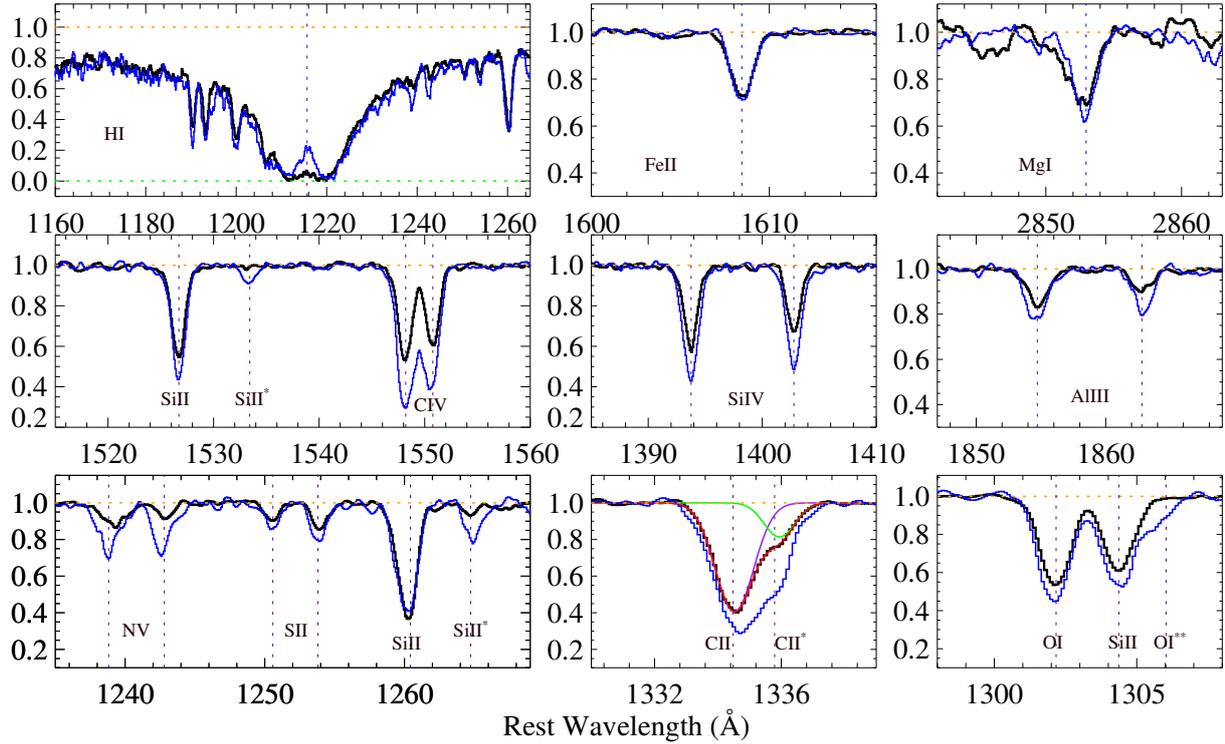}
\end{tabular}
\caption{Some important transitions detected in the composite spectra of eclipsing DLAs with weak (black curves) and strong (blue curves) narrow Ly$\alpha$ emission are shown. In the C\,{\sc ii} velocity panel, we decompose the absorption feature by conducting a double-Gaussian fit. Here, the purple (resp. green) curve represents the C\,{\sc ii} (resp. C\,{\sc ii}$^{*}$) component. The final fit is overplotted as a red curve.}
 \label{zoomedplots2}
\end{figure*}

The EWs of the species measured in the two composites are listed in Table\,\ref{ewtable} (columns\,5\,\&\,6), and Fig.\,\ref{zoomedplots2} presents the absorption lines of some important transitions. We also plot in Fig.\,\ref{EWvsEW} (lower panel), the EWs of absorption lines detected in one composite as a function of those detected in the other. As shown in Fig.\,\ref{EWvsEW}, the EWs of the low ionization species in the strong composite are either equal to or larger than those in the weak composite. This could be an indication that the metallicity is higher in DLAs with stronger narrow Ly$\alpha$ emission. Using \citet{2008ApJ...685..344P} calibration to estimate the metallicity from the Si\,{\sc ii}\,$\lambda$1526 EW, we get $-$1.15 and $-$1.0 from the weak and strong composites, respectively.

\begin{figure}
\centering
\begin{tabular}{c}
\includegraphics[bb=65 362 363 578, clip=,width=0.90\hsize]{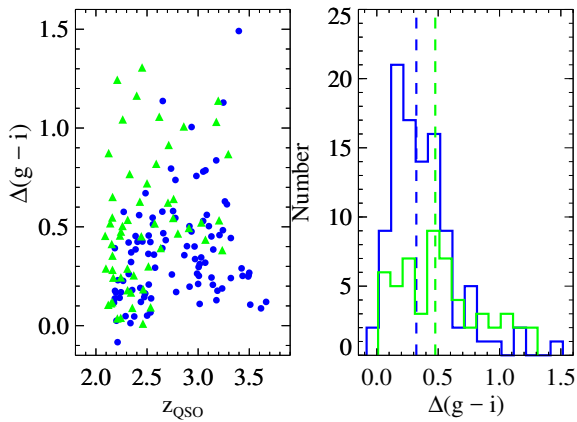}
\end{tabular}
\caption{$\Delta(g-i)$ color as a function of the quasar redshift (left panel), and the $\Delta(g-i)$ color distribution (right panel). Green data-points and histogram are for eclipsing DLAs with strong narrow Ly$\alpha$ emission while blue data-points and histogram represent eclipsing DLAs with weak narrow emission.}
 \label{g_minus_i}
\end{figure}

The 40 per cent higher C\,{\sc ii}$\lambda$1334/Fe\,{\sc ii}$\lambda$2344 EW ratio in the strong composite could be due to the higher level of depletion in the eclipsing DLAs with stronger emission. If eclipsing DLAs with stronger emission are dustier, then one would expect their foreground quasars to be redder. In order to explore the difference in the reddening of the quasar spectra from the two subsamples, we plot in Fig.\,\ref{g_minus_i} the $\Delta(g-i)$ color as a function of quasar redshift (left panel) along with the $\Delta(g-i)$ color distribution (right panel). Fig.\,\ref{g_minus_i} shows that quasars behind eclipsing DLAs with stronger narrow Ly$\alpha$ emission have redder average $\Delta(g-i)$ colors.

The difference in the EWs of the high ionization species (i.e. N\,{\sc v}, Si\,{\sc iv} and C\,{\sc iv}) in the two composites is more significant than what is seen in the low ionization species. For example, N\,{\sc v}\,$\lambda$1235 is more than a factor of 2 larger in the strong composite. This trend is also seen in the Al\,{\sc iii} absorption as it is a factor of 1.45 larger. Stronger absorption from the high ionization species implies that the DLAs with stronger narrow Ly$\alpha$ emission are exposed to a harsher radiation field. Since the median luminosity (at 1500\,\textup{\AA}) of the quasars from the strong subsample (i.e. $L_{1500}$\,=\,1.11\,$\times$\,10$^{42}$) is smaller than that of the weak subsample (i.e. $L_{1500}$\,=\,1.85\,$\times$\,10$^{42}$), this would imply that the DLAs from the strong subsample should be physically closer to the quasars.

Absorption from the fine structure states of Si\,{\sc ii}, C\,{\sc ii} and O\,{\sc i} are also detected in the two composites (Fig.\,\ref{zoomedplots2}). As shown in Fig.\,\ref{zoomedplots2}, absorption from C\,{\sc ii}$^{*}$\,$\lambda$1335 and Si\,{\sc ii}$^{*}$\,$\lambda$1264 are detected in both composites while O\,{\sc i}$^{**}$\,$\lambda$1306, Si\,{\sc ii}$^{*}$\,$\lambda$1309 and $\lambda$1533 are only detected in the strong composite. From Fig.\,\ref{zoomedplots2} and Table\,\ref{ewtable}, it is evident that fine structure lines are stronger in DLAs with stronger narrow Ly$\alpha$ emission.

Fine structure states can be populated by different processes: collisional excitation, UV pumping and direct excitation by the cosmic microwave background radiation (CMBR). Taking all these mechanisms into account, Silva and Viegas (2002) performed a detailed calculation of the theoretical population ratios of the fine structure states of several species such as Si\,{\sc ii}. From their figure\,8, one can see that at low density ($n_{\rm HI}$\,$<$\,1\,cm$^{-3}$), the observed Si\,{\sc ii}$^{*}$/Si\,{\sc ii} ratios (at T\,$\sim$\,10\,000\,K) imply that the DLAs in the strong subsample should be exposed to radiation fields that are stronger than the one the DLAs in the weak subsample are exposed to. On the other hand,  at higher densities (i.e. $n_{\rm HI}$\,$>$\,1\,cm$^{-3}$) the presence of a radiation field does not significantly affect the level populations. In this case, collisional excitation is the dominant process in populating the fine structure states. At higher densities, our observed Si\,{\sc ii}$^{*}$/Si\,{\sc ii} ratios would imply that the density of the DLAs in the strong subsample should be higher than the DLAs in the other subsample.

The stronger absorption from the high ionization species along with the higher ratio of the excited to the ground state of species like Si\,{\sc ii} suggest that the DLAs exhibiting stronger narrow Ly$\alpha$ emission should be denser and closer to the AGN.

\section{discussion and conclusion}

A similar study to the one presented here has been performed for 57 eclipsing DLAs from the SDSS DR9 and DR10 by \citet{2013A&A...558A.111F}. These author's efforts to distinguish between eclipsing DLAs with and without narrow Ly$\alpha$ emission were inconclusive, mainly due to the small sample size. However, in the current study, using a larger sample of eclipsing DLAs from the SDSS-III DR12, we could indeed distinguish three populations of such DLAs based on the strength of the narrow Ly$\alpha$ emission detected in the DLA trough: 1) DLAs that reveal no Ly$\alpha$ NEL in their troughs, 2) DLAs with weak Ly$\alpha$ NEL emission, and 3) DLAs with strong Ly$\alpha$ NEL. We recall that our statistical sample contains 399 eclipsing DLAs, out of which 155 systems reveal NEL in their DLA troughs.

By analyzing the composite spectra of these eclipsing DLAs, we found that DLAs with stronger NEL are also denser, and hence smaller in size. Smaller DLAs would then cover smaller fraction of the background extended Ly$\alpha$ emitting source. In this case, one would expect to detect stronger Ly$\alpha$ emission in the DLA trough. This could partly explain the correlation found between the strength of the NEL and the strength of the absorption from the excited states of species like Si\,{\sc ii}, O\,{\sc i}, and C\,{\sc ii}. Moreover, DLAs with stronger NEL also appear to be enveloped by more ionized layers of gas, as indicated by the strength of the absorption lines of high ionization species. Since quasars in the three subsamples have almost similar intrinsic emission properties, higher ionization in DLAs with stronger NEL implies that these clouds should be physically closer to the AGN. Since absorbers close to AGNs are expected to have near-solar or solar metallicity \citep{1994A&A...291...29P,2001A&A...373..816S}, the low metalicities of our eclipsing DLAs  suggest that these clouds could be part of some infalling material accreting on to the quasar host galaxy, perhaps through the filaments of the cosmic web.

One could think of these eclipsing DLAs as the product of the interaction between infalling and outflowing gas. The AGN-driven outflowing gas, on its way to the outskirt of the host galaxy, could collide with the infalling gas and compress it to a DLA \citep{2014MNRAS.443.2018N}. If the collision between the two gas clouds occurs closer to the quasar then the gas would become more compressed and denser. This could qualitatively explain why eclipsing DLAs with stronger NEL in their troughs are denser. Moreover, when the metal-poor infalling gas collide with the enriched outflowing gas, they get mixed, and the metallicity of the mixture of the two gas would be intermediate between those of the two colliding gas \citep{2017ASSL..430.....F}. Interestingly, the metallicities of our eclipsing DLAs (i.e. [X/H]\,$\sim$\,$-$1) seem to be consistent with what one would expect from the dilution of enriched outflowing gas by chemically young infalling gas. High resolution spectra of some of our best candidate eclipsing DLAs could allow us to resolve and trace the inflow and outflow absorption signatures in the spectra, and help in elucidating the nature of these DLAs.

\citet{2013A&A...558A.111F} proposed that eclipsing DLAs without NEL in their trough could be some neighboring galaxies near the quasars, which act as a screen, blocking all Ly$\alpha$ emission from the quasar host galaxy. Alternatively, it could be possible that these DLAs have the same origin as DLAs with NEL. It may well be that eclipsing DLAs that exhibit no narrow emission in their trough could also be tracing interaction between infalling and outflowing gas, but at a larger distance from the AGN. Detailed study of the chemical properties and ionization states of these DLAs would be needed to robustly confirm the validity of these scenarios.

\begin{table}
\caption{Rest equivalent widths for the four composites in mili-angstrom. First column: ID of the species. Second column: rest wavelengths. Third column: rest EWs for the composite of eclipsing DLAs with a NEL. Fourth column: rest EWs for the composite of eclipsing DLAs without a NEL. Fifth column: rest EWs for the composite of eclipsing DLAs with weak narrow Ly$\alpha$ emission. Sixth column: rest EWs for the composite of eclipsing DLAs with strong NEL.} 
\centering 
 \setlength{\tabcolsep}{5.0pt}
\renewcommand{\arraystretch}{1.05}
\begin{tabular}{c c c c c c} 
\hline\hline 
ID & $\lambda_{\rm lab}$ & compos.1 & compos.2 & compos.3 & compos.4 \\ [0.5ex] 
\hline 
N\,{\sc v} & 1238 & 142\,$\pm$\,12 & 234\,$\pm$\,6 & 235\,$\pm$\,16 & 483\,$\pm$\,15 \\
N\,{\sc v} & 1242 & 137\,$\pm$\,21 & 243\,$\pm$\,15 & 120\,$\pm$\,5 & 446\,$\pm$\,28 \\
Si\,{\sc iv} & 1393 & 582\,$\pm$\,6 & 734\,$\pm$\,8 & 585\,$\pm$\,13 & 953\,$\pm$\,18 \\
Si\,{\sc iv} & 1402 & 351\,$\pm$\,11 & 607\,$\pm$\,9 & 495\,$\pm$\,15 & 801\,$\pm$\,18 \\
C\,{\sc iv} & 1548 & 787\,$\pm$\,15 & 1018\,$\pm$\,10 & 795\,$\pm$\,22 & 1420\,$\pm$\,12 \\
C\,{\sc iv} & 1550 & 595\,$\pm$\,18 & 750\,$\pm$\,15 & 633\,$\pm$\,21 & 1189\,$\pm$\,18 \\
Al\,{\sc iii} & 1854 & 340\,$\pm$\,11 & 361\,$\pm$\,8 & 321\,$\pm$\,18 & 465\,$\pm$\,28 \\
Al\,{\sc iii} & 1862 & 211\,$\pm$\,7 & 301\,$\pm$\,16 & 201\,$\pm$\,24 & 370\,$\pm$\,23 \\
Al\,{\sc ii} & 1670 & 878\,$\pm$\,8 & 744\,$\pm$\,9 & 709\,$\pm$\,17 & 844\,$\pm$\,11 \\
S\,{\sc ii} & 1250 & 119\,$\pm$\,5 & 166\,$\pm$\,10 & 136\,$\pm$\,13 & 182\,$\pm$\,28 \\
S\,{\sc ii} & 1253 & 217\,$\pm$\,5 & 219\,$\pm$\,10 & 202\,$\pm$\,11 & 296\,$\pm$\,11 \\
Si\,{\sc ii} & 1260 & 1198\,$\pm$\,16 & 1065\,$\pm$\,27 & 1036\,$\pm$\,15 & 1156\,$\pm$\,19 \\
Si\,{\sc ii} & 1304 & 630\,$\pm$\,20 & 585\,$\pm$\,20 & 524\,$\pm$\,6 & 683\,$\pm$\,16 \\
Si\,{\sc ii} & 1526 & 909\,$\pm$\,14 & 765\,$\pm$\,12 & 692\,$\pm$\,18 & 890\,$\pm$\,19 \\
Si\,{\sc ii} & 1808 & 229\,$\pm$\,11 & 267\,$\pm$\,8 & 240\,$\pm$\,16 & 298\,$\pm$\,12 \\
Si\,{\sc ii}$^{*}$  & 1264 & 26\,$\pm$\,3 & 115\,$\pm$\,5 & 71\,$\pm$\,6 & 303\,$\pm$\,22 \\
Si\,{\sc ii}$^{*}$  & 1309 & ... & 47\,$\pm$\,8 & 33\,$\pm$\,13 & 113\,$\pm$\,16 \\
Si\,{\sc ii}$^{*}$  & 1533 & ... & 74\,$\pm$\,3 & 15\,$\pm$\,3 & 164\,$\pm$\,14 \\
Fe\,{\sc ii} & 1608 & 508\,$\pm$\,13 & 434\,$\pm$\,8 & 430\,$\pm$\,7 & 462\,$\pm$\,13 \\
Fe\,{\sc ii} & 2249 & 154\,$\pm$\,15 & 107\,$\pm$\,13 & ... & 98\,$\pm$\,11 \\
Fe\,{\sc ii} & 2260 & 158\,$\pm$\,13 & 129\,$\pm$\,5 & 82\,$\pm$\,14 & 153\,$\pm$\,66 \\
Fe\,{\sc ii} & 2344 & 1177\,$\pm$\,16 & 890\,$\pm$\,15 & 939\,$\pm$\,23 & 933\,$\pm$\,33 \\
Fe\,{\sc ii} & 2374 & ... & ... & 804\,$\pm$\,46 & 785\,$\pm$\,33 \\
Fe\,{\sc ii} & 2382 & 1342\,$\pm$\,11 & 1123\,$\pm$\,39 & 1051\,$\pm$\,71 & 1021\,$\pm$\,25 \\
Fe\,{\sc ii} & 2600 & 1300\,$\pm$\,16 & 1170\,$\pm$\,20 & 1080\,$\pm$\,22 & 1259\,$\pm$\,150 \\
Mg\,{\sc i} & 2852 & 510\,$\pm$\,15 & 663\,$\pm$\,21 & 729\,$\pm$\,10 & 766\,$\pm$\,16 \\
C\,{\sc ii} & 1334 & 1063\,$\pm$\,10 & 854\,$\pm$\,13 & 895\,$\pm$\,10 & 1220\,$\pm$\,15 \\
C\,{\sc ii}$^{*}$  & 1335 & 197\,$\pm$\,12 & 413\,$\pm$\,15 & 232\,$\pm$\,13 & 455\,$\pm$\,17 \\
O\,{\sc i} & 1302 & 826\,$\pm$\,15 & 658\,$\pm$\,15 & 623\,$\pm$\,17 & 762\,$\pm$\,18 \\
O\,{\sc i}$^{**}$  & 1306 & ... & ... & ... & 142\,$\pm$\,20 \\

\hline 
\end{tabular}
\label{ewtable} 
\end{table}

\section*{Acknowledgements}

We would like to thank the referee for carefully reading our paper and for giving constructive comments which helped improving the quality of the paper.
HFV would like to thank Jens-Kristian Krogager for useful discussion. PPJ, PN and RS
gratefully acknowledge the support of the Indo-French Centre for the promotion of Advanced Research
(IFCPR) under contract number 5504-2.
\noindent
Funding for SDSS-III has been provided by the Alfred P. Sloan Foundation, the Participating Institutions, the National Science Foundation, and the U.S. Department of Energy Office of Science. The SDSS-III web site is \url{http://www.sdss3.org/}.
\noindent
SDSS-III is managed by the Astrophysical Research Consortium for the Participating Institutions of the SDSS-III Collaboration including the University of Arizona, the Brazilian Participation Group, Brookhaven National Laboratory, Carnegie Mellon University, University of Florida, the French Participation Group, the German Participation Group, Harvard University, the Instituto de Astrofisica de Canarias, the Michigan State/Notre Dame/JINA Participation Group, Johns Hopkins University, Lawrence Berkeley National Laboratory, Max Planck Institute for Astrophysics, Max Planck Institute for Extraterrestrial Physics, New Mexico State University, New York University, Ohio State University, Pennsylvania State University, University of Portsmouth, Princeton University, the Spanish Participation Group, University of Tokyo, University of Utah, Vanderbilt University, University of Virginia, University of Washington, and Yale University.













\bibliographystyle{mn2e}
\bibliography{mn}

\end{document}